\documentclass[11pt]{article}

\usepackage[margin=1in]{geometry}
\usepackage[T1]{fontenc}
\usepackage[utf8]{inputenc}
\usepackage{amsmath}
\usepackage{amssymb}
\usepackage{caption}
\usepackage{subcaption}
\usepackage{graphicx}
\usepackage{tabularx}
\usepackage{booktabs}
\usepackage{hyperref}
\usepackage{natbib}
\usepackage{xcolor}
\usepackage{wrapfig}
\usepackage{placeins}
\usepackage{algorithm}
\usepackage{algpseudocode}
\usepackage{enumitem}
\usepackage{url}

\title{Using Theory of Mind to Arbitrate between Social and Non-social Learning}

\author{%
Lance Ying\textsuperscript{1,2,*},
Ryan Truong\textsuperscript{1,*},
Joshua B. Tenenbaum\textsuperscript{2},
Samuel J. Gershman\textsuperscript{1}%
\\[1.5em]
{\small\itshape \textsuperscript{1}Harvard University \quad
\textsuperscript{2}Massachusetts Institute of Technology}%
}

\date{}

\begin{document}

\maketitle

\renewcommand{\thefootnote}{\fnsymbol{footnote}}
\footnotetext[1]{L.Y. and R.T. contributed equally to this work.}
\renewcommand{\thefootnote}{\arabic{footnote}}
\setcounter{footnote}{0}

\begin{abstract}
Social learning is a powerful mechanism through which agents learn about the world from others. However, humans sometimes choose direct experience over social learning, which can carry time and cognitive resource costs. How do people balance social and non-social learning? We propose a Rational Mentalizing model of the decision to engage in social learning. This model estimates the utility of social learning by reasoning about  another agent's goal and the informativeness of their future actions. It then weighs the utility of social learning against the utility of non-social learning. Using a novel game where players choose between observing other agents or exploring the environment, we show that the Rational Mentalizing model can quantitatively capture human trade-offs between these strategies. These findings suggest that selective social learning is guided by `Theory of Mind' in the service of utility maximization.
\end{abstract}

\vspace{0.5em}

\section{Introduction}

Social learning---acquiring information through others---is a powerful mechanism by which humans and non-human animals acquire information about the world \citep{henrich2016secret, kendal2018social}. Yet humans do not use social learning indiscriminately. We choose \textit{when} to observe, \textit{how long} to watch, and \textit{whom} to learn from. Consider choosing a restaurant for dinner: you could browse on your own or follow where others go. But if other diners have different food preferences, following the crowd may not lead to the best outcome---many diners may be heading to a famous Italian restaurant while you are craving Asian food. To decide whether to follow others, you need to reason about whether their goals align with yours. This selectivity is central to the power of human social learning.

At the core of this reasoning is Theory of Mind (ToM), the ability to attribute mental states such as goals and beliefs to other agents \citep{wellman2004theory, wellman2002understanding}. Recent work in cognitive science has shown that ToM plays a crucial role in human teaching and learning \citep{gweon2021inferential, shafto2012learning, hawkins2023flexible, dutemple2023know}. These strategies enable learning from sparse, indirect data. Returning to the restaurant example, you might infer that your friend knows a good restaurant but is not hungry, or is hungry but does not know where to go; only by reasoning about both her knowledge and her goals can you decide whether to follow her. You could also observe your knowledgeable friend when she is hungry so that you can use this information later when you are hungry yourself. In each case, the learner is not copying behavior but reasoning about the mental states that generated it.

Existing computational accounts of social learning, however, bypass this generative view of the observed agent. Heuristic accounts characterize social learning in terms of simple policies; copy the majority, copy the successful, copy the prestigious \citep{laland2004social, rendell2011cognitive, muthukrishna2016and, heyes2012s}, and have been productive in capturing regularities in animal and cross-cultural data, as well as in establishing that social learners enjoy efficiency advantages over purely individual learners \citep{kendal2005trade, henrich2016secret}. Reinforcement learning accounts go further: rather than applying fixed copying rules, they let learners weight social information against individual information by each source's estimated reliability, inferred from how volatile that source has been over time \citep{behrens2008associative, diaconescu2020neural}. This logic has since been extended to advice-taking, norm learning, and feature-based reward learning \citep{biele2009computational, pereg2024disentangling, hertz2021learning, schultner2025feature}. Yet both heuristic and reinforcement learning accounts share a blind spot: they treat observed behavior as a signal to be weighted, rather than as evidence about the goals and beliefs that produced it. As a consequence, they struggle to explain how humans learn efficiently from sparse, indirect observations, where the relevant inference is not \textit{what} the other agent did but \textit{why} they did it \citep{velez2021learning, gweon2021inferential, bonawitz2016computational}.

Employing Theory of Mind for social learning comes with a cognitive cost relative to non-social learning. Thus, people may opt for non-social learning when these costs are too high relative to their expected benefits. Consistent with this hypothesis, \citet{kendal2005trade} demonstrated that people are more likely to engage in social learning when the cost of non-social learning increases. More recently, \citet{wu2025adaptive} showed, in an immersive collective foraging task, that humans adaptively modulate their reliance on social versus non-social cues as a function of environmental structure and recent foraging success. These and other studies focused on tasks where all agents pursue the same goal, so another agent's success is by construction informative about the observer's own rewards. In a collective sensing task used by \citet{hawkins2023flexible}, for instance, once the observer can infer who has knowledge of the goal location, social learning tends to be less costly than non-social learning. In more realistic settings where agents have different goals, the value of observing another agent depends on whether their behavior is informative for the observer's own objectives.

Outside social learning specifically, a parallel literature on active information seeking has shown that humans rationally select queries based on expected information gain, balancing the value of new evidence against the cost of obtaining it \citep{oaksford1994rational, markant2014select, coenen2019asking, ruggeri2019shake}. This framework applied cleanly to non-social tasks-choosing which cards to flip, which experiments to run, and which hypotheses to test, but its extension to social sources is not simple. When the source is another agent, the informativeness of an observation depends on the agent's goals and beliefs, not on a fixed query-answer mapping. A rational social learning agent must therefore combine the cost-benefit logic of active learning with a generative model of the agent being observed.

We make two contributions. First, we propose a ``Rational Mentalizing'' model of social learning, grounded in the Bayesian Theory of Mind (BToM) framework \citep{baker2017rational, jara2019theory, ying2025language}. BToM models other agents as rational planners and infers their goals and beliefs from observed actions \citep{baker2009action, zhixuan2024pragmatic, ying2024grounding, stacy2021modeling}. Our model extends this framework: the observer uses BToM not merely to understand another agent, but to estimate whether that understanding is worth the cost of obtaining it. The model weighs the expected utility of social learning---computed by mentalizing about other agents' goals and plans---against the utility of non-social exploration, to decide if, when, and whom to observe. Second, we introduce a novel experimental paradigm in which participants navigate a multi-agent environment and must decide on each step whether to observe another agent or act independently. The game is designed to create rich, parametrically controllable trade-offs between social and non-social learning across varying levels of goal uncertainty, agent availability, and agent expertise. Across four experiments, we systematically vary the environment to create increasingly complex trade-offs.

\section{Social Learning Game}

\begin{figure*}[t]
    \centering
    \includegraphics[width=\textwidth]{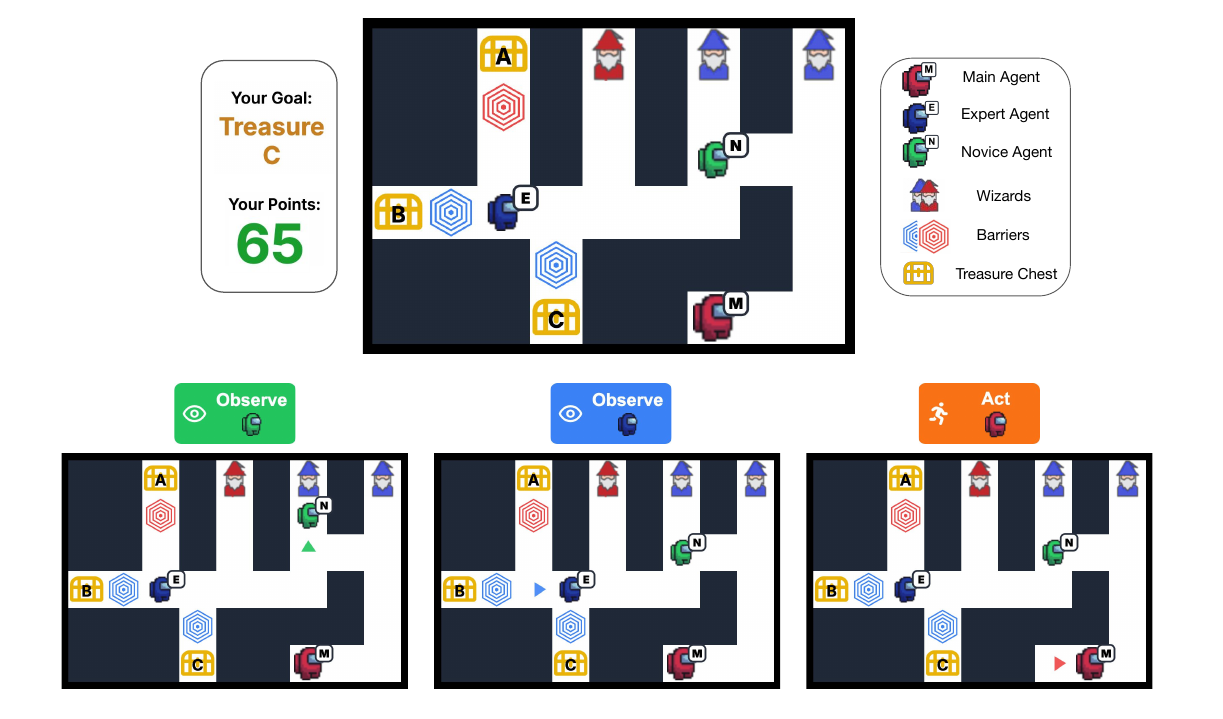}
    \caption{Illustration of the social learning game. The participant controls the red (main) agent and must retrieve a target treasure chest. On each timestep, the participant can either observe a selected NPC (causing only that NPC to move) or act (moving the red agent one step). The map contains wizards (only one of each color holds the amulet), colored barriers, and treasure chests. The legend shows the main agent, expert NPCs, wizards, barriers, and treasure chests.}
    \label{fig:exp_fig}
\end{figure*}

The game takes place on a 12 x 11 grid map containing treasure chests, wizards, and colored barriers (Figure~\ref{fig:exp_fig}). Each treasure chest is either unobstructed or blocked by a colored barrier (e.g., red or blue). To pass through a barrier, a player must obtain an amulet of the matching color from a wizard. Among all wizards of a given color, only one holds the amulet while the rest are decoys. The number and placement of wizards varies across maps.

In every experiment, a red agent is controlled by the human participant to reach one of the treasure chests. The red agent does not know which wizard has the amulet. In addition to the red agent, one or more other agents agents (non-player characters, or NPCs) are also present on the map. All agents can see each other's positions, but each agent's actions do not affect any other agent's game state. Obtaining an amulet, passing through a barrier, or retrieving a chest has no effect on other agents' ability to do the same. Each agent has a goal of retrieving a specific lettered treasure chest, but agents do not know each other's goals. By default, NPCs know which wizard among a set of same-colored wizards holds the amulet (``expert'' expertise). However, in one experimental condition (Experiment 4), one of the two observable NPCs lacks full knowledge of the map structure (``novice'' expertise).

At each timestep, the red player agent can either observe a chosen NPC or move one step in one of four cardinal directions: up, down, left, or right. If the agent chooses to observe, only the selected NPC moves one step while all other agents remain still. If the agent chooses to move, only the red agent moves and all other agents remain still. The red agent can also observe whenever an NPC interacts with a wizard, a barrier, or a treasure chest. Since the red player agent does not know which wizards possess amulets, it must weigh whether the information gained through observation is worth the cost of exploring on its own.

We designed four experiments, each conducted as a separate study with distinct participants, that systematically vary the number of observable agents, treasure chests, and expertise levels (see Methods for full details). Experiment 1 uses a single expert NPC and a single treasure chest, isolating the basic trade-off between observing and exploring. Experiment 2 adds two additional treasure chests, introducing goal uncertainty. Experiment 3 adds a second expert NPC, requiring participants to choose \textit{whom} to observe. Experiment 4 replaces one expert with a novice NPC, testing whether participants calibrate observation to agent expertise accordingly.

\section{Rational Mentalizing Model}

In this paper we propose a rational mentalizing model of social learning. The model performs a simple rational utility comparison at each timestep: \textit{is any information I gain by watching another agent worth more than what I would lose by not acting on my own goal?} To do this, it estimates the utility of observing and acting, and choose the action with the higher estimated utility.

First, $C_{\text{self}}(b_t)$, is the expected cost of committing now and executing the observer's best plan under its current beliefs about the environment (e.g., which of the remaining wizards holds the amulet. The second, $Q_{\text{obs}}^j(t)$, is the expected cost of watching agent $j$ for some number of steps and \textit{then} acting. Computing $Q_{\text{obs}}^j(t)$ requires the observer to imagine, for each plausible hypothesis about what the watched agent wants and believes, how that agent's future trajectory would refine the observer's own beliefs and therefore reduce the cost of the observer's eventual plan.

\begin{figure}[!htbp]
    \centering
    \includegraphics[width=0.5\textwidth]{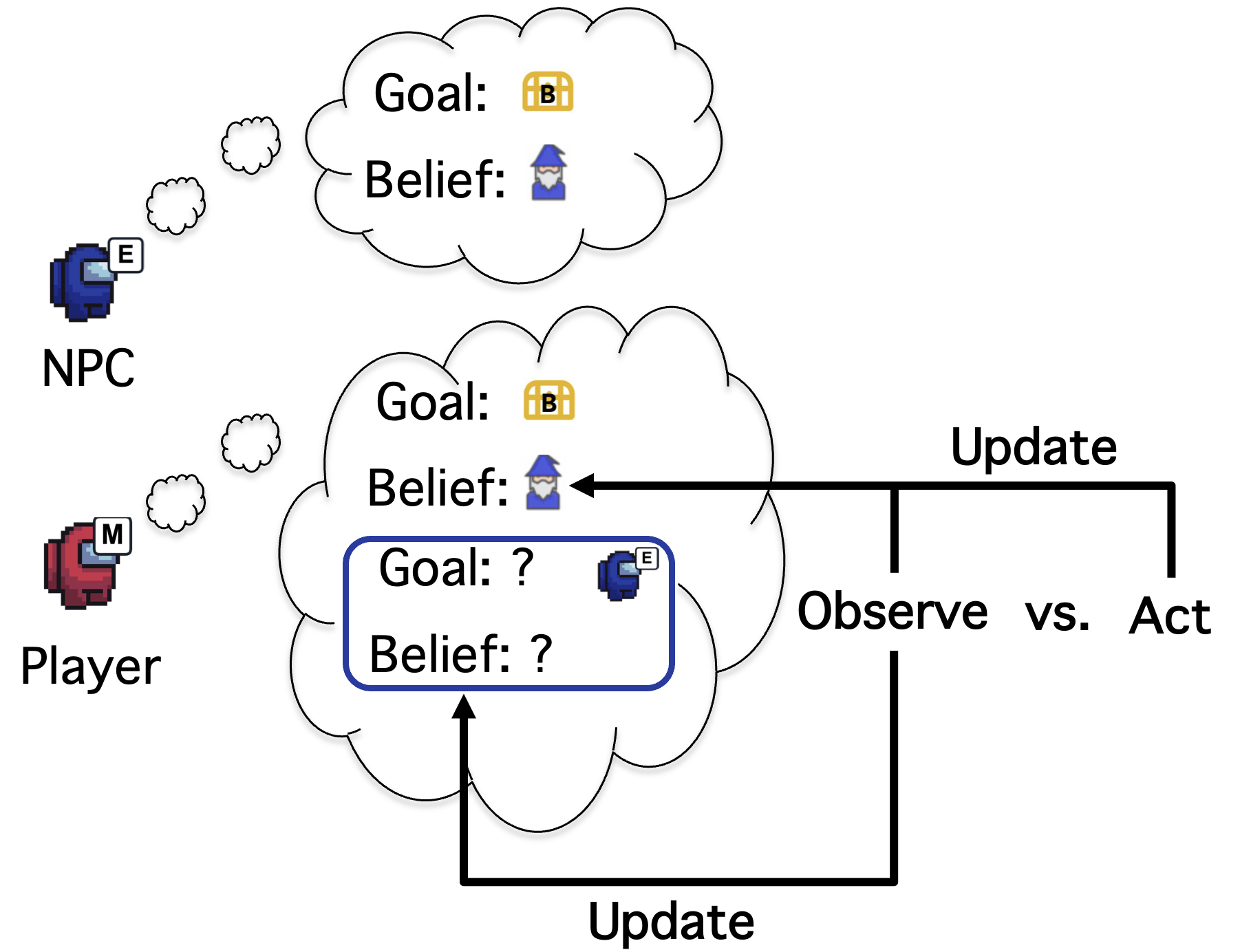}
    \caption{\textbf{Schematic of the Rational Mentalizing model.} At each timestep, the observer (red agent) decides whether to act or observe another agent. \textbf{Utility comparison:} it weights the cost of acting now ($C_{\text{self}}$) against the expected cost of watching the expert for another step ($Q_{\text{obs}}$), and observes only when watching is cheaper. \textbf{Theory of Mind:} to establish that cost, the observer infers the observed agent's goal and belief, for example whether it is heading to treasure A or B, and predicts how much observing would reduce its own uncertainty.}
    \label{fig:model}
\end{figure}

Three ingredients make this possible (Figure~\ref{fig:model}): Bayesian inference over the observed agent's goal and belief; simulation of how that agent's future actions would narrow the observer's candidate set; and a cost comparison between the resulting expected plan cost (plus time spent watching) and the cost of acting now. Two properties of this formulation matter for what follows. First, the observer does not need to know the other agent's goal or belief, it only needs a posterior over them, which is updated by every action it observes. Second, the model naturally predicts \textit{whom} to watch when multiple agents are available: the choice falls out of which agent's hypothesized trajectory is expected to reduce the observer's candidate set the most, per unit of observation steps. We now formalize these ideas.

\begin{algorithm}
\caption{Rational Mentalizing Model}
\label{alg:rational_mentalizing}
\begin{algorithmic}[1]
\Require Observer with beliefs $b_t$; observable agents $\{j\}$; particles $\{(\theta_i^{j}, p_t^{i})\}_{i=1}^N$ for each agent $j$

\While{True}
    \State Compute non-social learning cost: $C_{\text{self}}(b_t)$
    \For{each observable agent $j$}
        \For{each particle $i \in \{1, \dots, N\}$}
            \State Simulate agent $j$'s future trajectory $\hat{a}_{t:t+T}^{j}$ under hypothesis $\theta_i^{j}$
            \State Compute updated beliefs $b_t'$ from simulated trajectory
            \State Estimate net cost: $q^{i} = c_{\text{obs}} T^{i} + C_{\text{plan}}(b_t')$
        \EndFor
        \State Aggregate: $Q_{\text{obs}}^j(t) = \sum_i p_t^{i} q^{i}$
    \EndFor
    \If{$C_{\text{self}}(b_t) \leq Q_{\text{obs}}^j(t)$ for all $j$}
        \State \textbf{Stop observing}
    \Else
        \State Observe $j^* = \arg\min_j Q_{\text{obs}}^j(t)$
        \State Update posterior $P(\theta^{j^*} \mid a_{1:t}^{j^*})$ via Bayesian inverse planning
    \EndIf
\EndWhile
\State \textbf{Act:} Execute belief-space planner using final posterior $P(g, b \mid a_{1:t})$
\end{algorithmic}
\end{algorithm}

\paragraph{Cost Comparison}
At step $t$ , the observer compares acting now against continuing to watch. Fix one hypothesis $\theta_i^{j} = (g_i^{j}, b_i^{j})$ (particle $i$) about the observed agent $j$'s goal and belief (for an expert NPC, the agent's belief coincides with the true amulet location), with posterior probability $p_t^i$. If that hypothesis were true, the observer can simulate forward and read off two things: $T^{i}$, how many more steps of watching it would take before the agent's trajectory narrows the remaining wizard to one (or reveals the hypothesis is wrong); and $C_{\text{plan}}^i$, the exploration cost the observer would still face afterward, over whatever candidates remain. The observer does not know which hypothesis is correct, so it marginalizes over its uncertainty:
\begin{align}
    Q_{\text{obs}}^j(t) = \sum_i p_t^{i} \left[ c_{\text{obs}} T^{i} + C_{\text{plan}}^{i} \right].
\end{align}
Here $c_{\text{obs}}$ is the per-step cost of observing. Let $j^* = \arg\min_j Q_{\text{obs}}^j(t)$. If $Q_{\text{obs}}^{j^*}(t) < C_{\text{self}}(b_t)$, the observer watches agent $j^*$ for one step, updates on the realized action, and re-evaluates; otherwise it commits and acts. Observation pays off when the hypotheses disagree about how quickly the candidate set would be narrowed and what would remain uncertain afterward.

\paragraph{Simulation and Belief Update}

Equation~(1) leaves three quantities unspecified: the posterior weights $p_t^i$ and, under each hypothesis, the observation horizon $T^i$ and the residual planning cost $C_{\text{plan}}^i$. All three derive from a single generative model of the observed agent,a near-optimal planner whose actions reveal its goal and belief. We first describe how this model yields the posterior $p_t^i$ (Eqs.~2--3), then how rolling it forward yields $T^i$ and $C_{\text{plan}}^i$.

We model each observed agent $j$ as a near-optimal planner in a partially observable environment \citep{zhi2020online, baker2017rational}. The observer does not know the other agent's goal $g^j$ (which treasure chest) or beliefs $b^j$ (which wizard holds the amulet). Under each goal-belief hypothesis $\theta^j = (g^j, b^j)$, the observed agent selects actions according to a Boltzmann policy over the cost-to-go $Q^*(s, a \mid \theta^j)$ of an optimal planner:
\begin{align}
    P(a_t^j \mid s_t, \theta^j) = \frac{\exp(-\beta \, Q^*(s_t, a_t \mid \theta^j))}{\sum_{a^\prime} \exp(-\beta \, Q^*(s_t, a^\prime \mid \theta^j))}.
\end{align}
where $\beta$ is a temperature parameter controlling stochasticity. This soft-optimality assumption assigns non-zero likelihood to suboptimal actions, enabling Bayesian inference from noisy observations. Given a sequence of observed actions $a_{1:t}^j$, the observer infers a joint posterior over the other agent's goal and beliefs by inverting the forward model via Bayes' rule:
\begin{align}
    P(\theta^j \mid a_{1:t}^j) \propto \prod_{\tau=1}^{t} P(a_\tau^j \mid s_\tau, \theta^j) \, P(\theta^j).
\end{align}
Each hypothesis $\theta^j$ is scored by how well it explains the observed trajectory. Critically, the observer does not have access to $j$'s actual beliefs; $b^j$ is a latent variable inferred jointly with $j$'s goal. We represent the posterior using a particle filter derived from the Sequential Inverse Plan Search (SIPS) algorithm \citep{zhi2020online}, which maintains a weighted set of particles $\{(\theta_i^{j}, p_t^{i})\}_{i=1}^N$ and updates their weights with each observed action. These posteriors are precomputed for all observable trajectories and cached, then read during the decision process rather than recomputed online.

These equations close the loop with Eq.~(1). The posterior $P(\theta^j \mid a_{1:t}^j)$ supplies the weights $p_t^{i}$. To obtain $T^{i}$ and $C_{\text{plan}}^{i}$, the observer rolls each particle $\theta_i^{j}$ forward under the same Boltzmann policy (Eq.~2), generating a simulated trajectory $\hat{a}_{t:t+T}^{j}$: $T^{i}$ is the step at which this trajectory first narrows the observer's candidate set to one (or contradicts the hypothesis), and $C_{\text{plan}}^{i} = C_{\text{plan}}(b_t')$ is the observer's exploration cost under the updated belief $b_t'$ that incorporates the simulated trajectory. When the observer instead chooses to observe, it updates the posterior using the actual observed action and re-evaluates Eq.~(1), and the decision process repeats. The full procedure is summarized in Algorithm~\ref{alg:rational_mentalizing}. Model parameters are fixed across experiments; values and sensitivity analyses are reported in the SI.

\section{Results}
We compared the full Rational Mentalizing model against three ablations, each removing one component of the full model. The \textbf{Rational Observer} weighs observation cost against exploration cost without reasoning about other's goals or beliefs. The \textbf{Mentalizing Observer} reasons about other agents' mental states but does not weigh social against non-social learning utility. The \textbf{Naive Observer} observes other agents indiscriminately, with no mentalizing or cost-benefit reasoning.

\subsection{Observation Behavior}
\begin{figure*}[t]
    \centering
    \includegraphics[width=\textwidth]{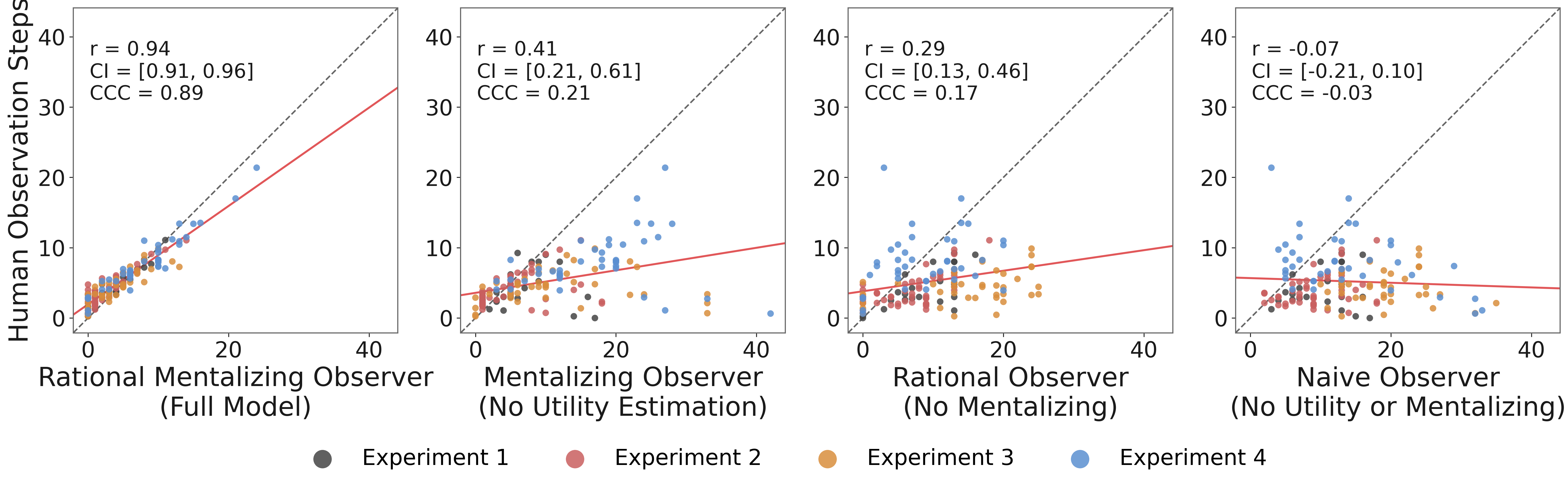}
    \caption{Observation steps pooled across all four experiments. Each panel shows one model, with points colored by experiment. Pearson $r$ and Lin's concordance correlation coefficient (CCC) with 95\% bootstrap confidence intervals are shown. The Rational Mentalizing model achieves the highest agreement with human observation behavior (CCC $= 0.89$), while the alternative models show low concordance, indicating poor absolute agreement.}
    \label{fig:obs_scatter}
\end{figure*}

Across four experiments of progressively increasing complexity, the Rational Mentalizing model captured observation behavior with high fidelity (per-experiment $r = 0.90$-$0.95$; pooled CCC $= 0.89$), while three ablations that each removed one component fitting substantially worse (Figure \ref{fig:obs_scatter}). The Mentalizing Observer (no utility estimation) tracked human behavior directionally but systematically over-observed. The Rational Observer (no mentalizing) and Naive Observer (neither) showed low concordance with human observation choices, particularly as task complexity increased.

Human observation behavior was internally consistent across the four experiments. Split-half reliability of observation counts ranged from $r_{SB} = 0.83$ to $r_{SB} = 0.92$ across experiments (Spearman-Brown corrected over 1000 random participant splits), establishing an upper bound on the fit any model could achieve against this data. We compared the four models against this ceiling on two metrics: the number of observation steps participants did per trial, and the total step counts (which depend on observation choices) serving as a downstream check.

\begin{figure*}[!p]
    \centering
    \includegraphics[width=\textwidth]{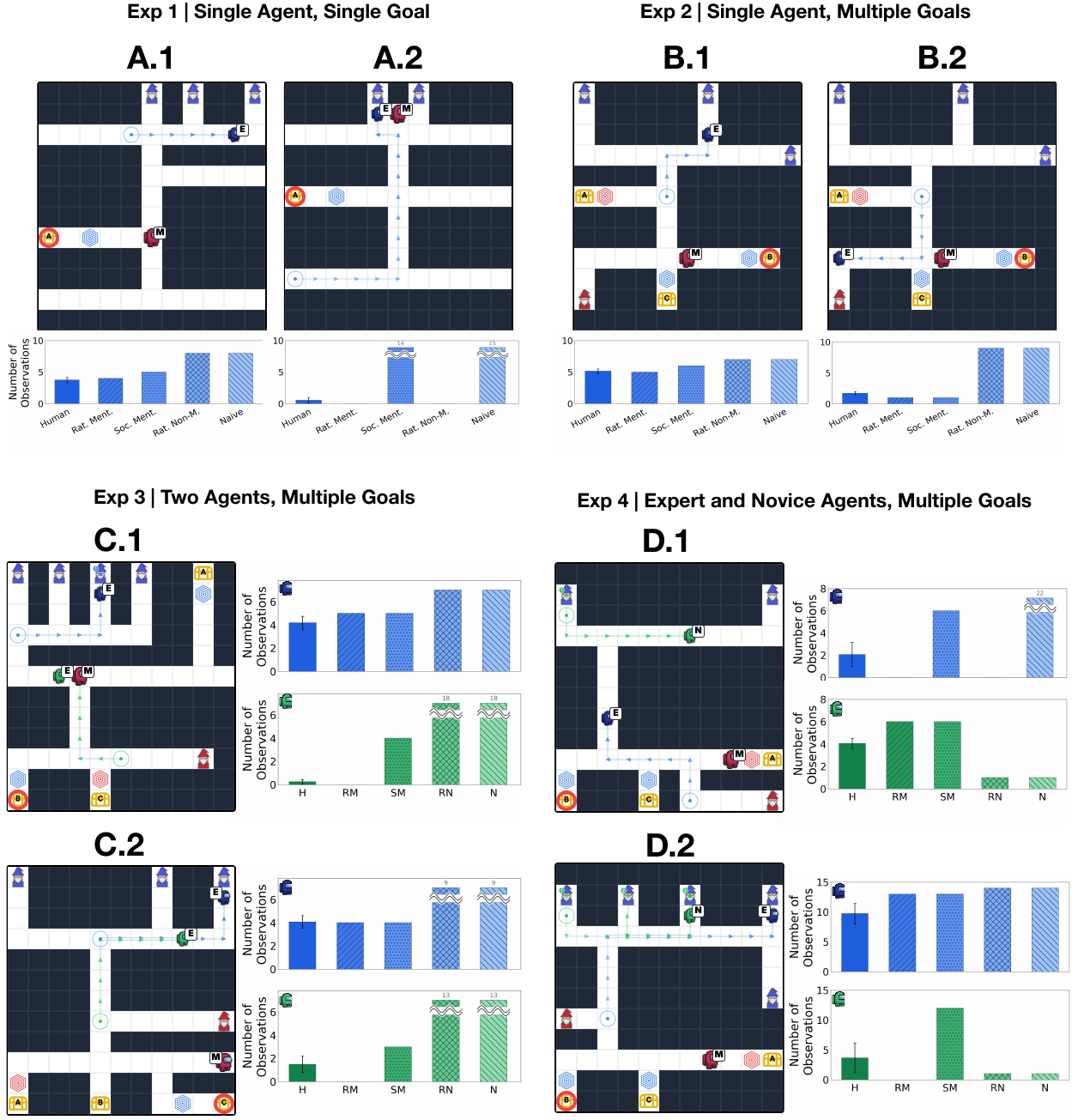}
    \caption{Qualitative examples of observation behavior across the four experiments. Each panel shows a trial map with agent trajectories and a bar chart comparing the number of observation steps for humans (H) and four models: Rational Mentalizing (RM), Mentalizing Observer (SM), Rational Observer (RN), and Naive Observer (N). Panels A.1--A.2: Experiment 1 (single agent, single goal). Panels B.1--B.2: Experiment 2 (single agent, multiple goals). Panels C.1--C.2: Experiment 3 (two expert agents, multiple goals), with separate observation counts for Agent 2 (blue) and Agent 3 (green). Panels D.1--D.2: Experiment 4 (expert and novice agents, multiple goals). Error bars indicate standard error across participants.}
    \label{fig:exp_results_qual}
\end{figure*}

In Experiment 1 (single expert agent, single goal), the Rational Mentalizing model's observation predictions correlated strongly with human behavior ($r = 0.94$, CI $[0.86, 0.98]$). The three alternatives showed lower agreement (Rational Observer $r = 0.66$, Naive Observer $r = 0.16$, Mentalizing Observer $r = 0.06)$. The Mentalizing Observer's correlation reflects its missing component: in three maps where the main agent began close to the available wizards, self-exploration was cheap relative to observation, and human and the Rational Mentalizing model accordingly skipped observation almost entirely (Figure~\ref{fig:exp_results_qual}, panel A.2) while the Mentalizing Observer continued to observe because it has no mechanism to weigh the observation cost against self-exploration. The same dissociation holds in panel A.1: when the expert's path could quickly reveal the amulet, humans and the Rational Mentalizing model observed for only a few steps before acting independently, while the Rational and Naive Observers watched the full trajectory and the Mentalizing Observer over-observed by one step.

In Experiment 2 (single expert agent, three goals - introducing goal uncertainty), the Rational Mentalizing model maintained strong agreement with human behavior ($r = 0.91$, CI $[0.81, 0.96]$). The three alternatives all fell well short (Rational Observer $r = 0.67$, Mentalizing Observer $r = 0.41$, CI $[0.10, 0.73]$, Naive Observer $r = 0.37$). Goal uncertainty is now central: the expert's trajectory carries information about both the amulet's locations and the agent's intended chest, and the observer must reason about both to decide whether continued watching is worthwhile. The Mentalizing Observer can do this reasoning but cannot weigh expected information gain against observation cost; the non-mentalizing alternatives cannot do it at all. Panels B.1 and B.2 illustrate the consequence: when the expert's early movements were diagnostic of both its goal and the amulet location, the Rational Mentalizing model, like humans, observed briefly; when those movements were ambiguous, both observed longer, while the Rational and Naive Observers observed for a fixed duration regardless and the Mentalizing Observer over-observed by one step in B.1.

\begin{figure*}[t]
    \centering
    \includegraphics[width=\textwidth]{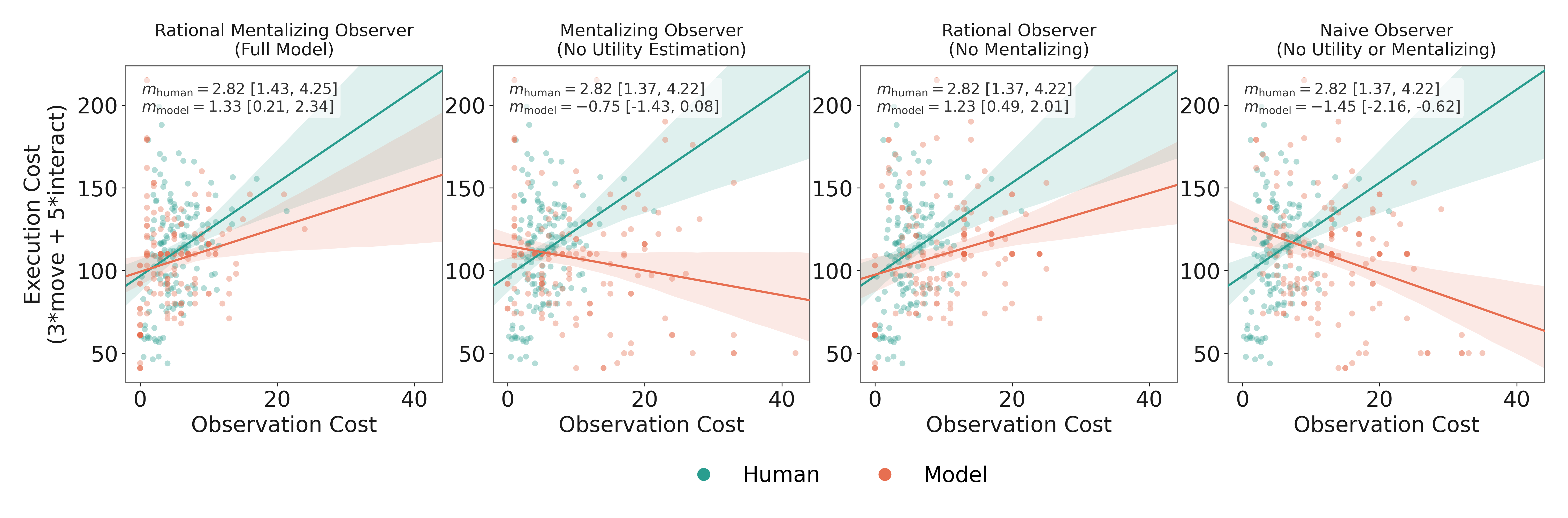}
    \caption{Observation cost vs.\ execution cost (3$\times$move + 5$\times$interact), pooled across experiments; teal = human means, coral = model, with each panel's fitted slope and its 95\% bootstrap CI band. Humans show a positive coupling---more observation accompanies more execution. The Mentalizing and Naive Observers invert this relationship (negative slopes), whereas the full model and the Rational Observer recover positive slopes. Among these, the full model's fit and confidence band overlap the human band most closely, the best qualitative match.}
    \label{fig:cost_scatter}
\end{figure*}

Experiment 3 involved two expert agents and three goals, allowing us to evaluate how humans and models determined whom to observe. The Rational Mentalizing model captured human observation patterns for both agents (agent 2: $r = 0.90$, CI $[0.86, 0.96]$; agent 3: $r = 0.93$, CI $[0.87, 0.96]$). All three alternatives fell short of the full model (Mentalizing Observer agent 2: $r = 0.47$, agent 3: $r = 0.27$; Rational Observer agent 2: $r = 0.61$, agent 3: $r = 0.36$; Naive Observer agent 2: $r = 0.40$, agent 3: $r = 0.25$). This drop is the signature of the new inference introduced in this experiment: the observer must compare which of two agents' future trajectories is expected to be more informative, weighed against the cost of watching either. The Naive and Rational Observers cannot make this comparison because they have no representation of the agents' goals; the Mentalizing Observer can identify which agent might be more informative directionally but lacks the utility comparison to allocate observation accordingly. When one agent's position made its future trajectory more informative about the amulet relevant to the observer's goal, humans and the Rational Mentalizing model concentrated observation on that agent while minimizing watching of the other, the Rational and Naive Observers observed both for a fixed duration, and the Mentalizing Observer over-observed the less-informative agent (Figure \ref{fig:exp_results_qual}).

Experiment 4 involved one expert and one novice agent, with three goals, allowing us to evaluate how humans and models calibrated their observing to varying expertise. The Rational Mentalizing model captured human observation of both agents (expert: $r = 0.90$, CI $[0.72, 0.96]$; novice: $r = 0.95$, CI $[0.91, 0.97]$). All three alternatives fit far worse than the full model (Mentalizing Observer expert: $r = 0.25$, novice: $r = 0.23$; Rational Observer expert: $r = 0.57$, novice: $r = 0.33$; Naive Observer expert: $r = 0.24$, novice: $r = -0.41$). Experiment 4 requires reasoning not only about an agent's goal but also about how that agent updates its beliefs as it interacts with the environment. The Mentalizing Observer has the inferential engine to do this but no utility comparison to know when the resulting belief shift is worth the watching cost. The Rational and Naive Observers instead default (mostly) to the expert, a strategy that succeeds when the expert is in fact the more informative source, and fails when the novice's behavior carries more counterfactual signal (Figure \ref{fig:exp_results_qual}, D.1-D.2). In panel D.1 (sparse wizards, novice positioned near one), the novice's continued trajectory after a wizard interaction reveals the amulets location in few observations, and humans and the Rational Mentalizing model accordingly observe the novice while alternatives default to the expert. In panel D.2 (many wizards, novice positioned closely to one of them), the same cue eliminates only a small fraction of hypotheses, and observation shifts to the expert; here the Mentalizing Observer flips and over-observes the novice, while the Rational and Naive Observers coincide with the Rational Mentalizing model by defaulting to the expert by policy.

Pooled across all experiments, the Rational Mentalizing model achieved Lin's concordance correlational coefficient $\text{CCC} = 0.89$ (CI $[0.84, 0.92]$) on observation counts, while all three alternatives showed low concordance ($\text{CCC} \leq 0.21$); the alternatives' 95\% bootstrap CIs do not overlap with the Rational Mentalizing model's, showing reliable differences in absolute agreement. The Rational Mentalizing model's CCC approaches the upper bound implied by human inter-participant reliability ($r_{SB} = 0.83$-$0.92$), leaving little remaining systematic variance for any model to capture.

The same pattern holds for the relationship between observation and execution (Figure~\ref{fig:cost_scatter}). For humans, levels with more observation also involve more execution, a positive slope that only the Rational Mentalizing model reproduces. The Rational Observer recovers a weaker positive slope whose confidence interval includes zero, while the Mentalizing and Naive Observers invert it with negative slopes.

\section{Discussion}

We proposed and tested a Rational Mentalizing model of adaptive social learning. The model integrates two capacities, Bayesian Theory of Mind and rational cost-benefit analysis, that have been studied largely in isolation, into a unified framework for deciding whether, when, and whom to observe. Across four experiments that systematically varied the complexity of the social learning environment, the model quantitatively captured human observation behavior, substantially outperforming three alternative models that each isolated on of it core components.

The ablation analysis reveals that both components of the model are necessary, and that they fail in opposite directions when removed. The Mentalizing Observer model, which reasons about other agent's mental states but does not weigh social learning against self-exploration, tends to over-observe: it continues observing agents even when the expected information gain does not justify the cost. The Rational Observer model, which performs cost-benefit analysis but does not reason about agents' goals, cannot anticipate whether future observations will be informative, leading to under-observing when the immediate payoff is unclear. The Naive Observer fails on a different question, the whom decision introduced in Experiments 3 and 4, where indiscriminate observation cannot recover. These patterns were especially pronounced in the observation count metric, which proved more diagnostic than total planning cost for distinguishing between models, a finding consistent with the idea that the decision of \textit{how much} to observe, rather than overall task performance, is where human social learning is most distinctively rational.

RL accounts of social learning have shown that humans dynamically arbitrate between social and individual information sources as a function of their relative reliability \citep{behrens2008associative, diaconescu2020neural, biele2009computational, pereg2024disentangling, hertz2021learning, schultner2025feature}. Our results suggest that this arbitration is not based solely on signal reliability: human observers also adjust observation as a function of the inferred goal and beliefs of the source, observing more when another agent's trajectory is expected to be informative for their own objectives and less when it is not. Capturing this requires representing the observed agent as a planning system, not as a noisy reward channel \citep{velez2021learning, gweon2021inferential}

Heuristic accounts of social learning, which characterize observation as following fixed rules such as ``copy the majority'' or ``copy the successful'' \citep{laland2004social, rendell2011cognitive, muthukrishna2016and}, similarly cannot accommodate the behavior we see here. A sharper comparison comes from utility based accounts developed in the same goal environment \citet{kendal2005trade}, \citet{hawkins2023flexible}, and \citet{wu2025adaptive} show that humans module reliance on social cues with foraging success and environmental structure, but in those tasks another agent's success is by construction informative about the observer's own rewards. Once goals diverge, as in Experiments 2--4, the value of observation cannot be read off the other agent's outcomes; it must be computed by reasoning about whose goals make whose actions informative. The closest empirical precedent is \citet{velez2019integrating}, who showed that learners integrate advice with private knowledge by inferring the process that generated it-recovering how an advisor selects advice rather than merely tracking its reliability. We share this posterior-inference logic but apply it where their paradigm doesn't reach: the observer learns from another agent's costly, goal-directed actions-not cooperative advice about a shared payoff-within a sequential decision problem, so that what the source's behavior reveals depends on first inferring goals that may differ from the observer's own.

A separate strand of cognitive science has examined how learners selectively trust some sources over others. Children and adults preferentially learn from sources judged to be more accurate, knowledgeable, or expert \citep{koenig2005preschoolers, harris2012trusting, sperber2010epistemic}, with this selectivity sensitive to source cues such as accuracy track records and explicit knowledge attribution. Experiment 4 suggests that humans calibrate observation not to expertise as a static property of an agent but to the expected informativeness of that agent's behavior in the current environment: across two map structures, the agent humans preferred to observe reversed. In our model, this calibration is not a separate trust mechanism but a consequence of mentalizing. An agent whose belief state is uncertain produces actions whose informativeness is itself a function of that uncertainty, and the rational arbitration absorbs this directly into the cost of observation. Selective trust and selective observation may then be products of a unified Bayesian inference over agent models rather than distinct cognitive systems.

The Rational Mentalizing model closely captures the qualitative pattern of human observation, tracking how often people choose to observe across conditions, but it underpredicts how much they do so. The same holds for total steps taken: the model reproduces the relative pattern across conditions while systematically underpredicting the absolute amount across all four experiments. In both cases, humans observe and move more than the model predicts. The gap is concentrated in scenarios where the model commits cleanly to a single information source while humans maintain a small but non-zero rate of observation of the alternative agent. This residual observation suggests participants hedge beyond what the rational arbitration prescribes, perhaps reflecting uncertainty in their own model of other agents, a softer-than-argmax decision, or curiosity-driven monitoring.

Three further questions remain open. First, our model treats observation as the only social channel; real social learning includes communication, demonstration, and pedagogical intent, where the observed agent optimized their actions for the observer rather than for their own goal \citep{shafto2014rational, gweon2021inferential, bonawitz2011double, bridgers2020young}. Extending the framework to those channels is not trivial. Second, the model treats each trial as independent and does not represent how humans accumulate agent-specific priors across repeated interactions. In real social learning, observers accumulate histories of which agents tend to be informative or trustworthy, and these histories shape future observation choices \citep{koenig2005preschoolers, harris2012trusting}. Third, the model captures how observation varies across conditions but consistently predicts less of it than people actually do: across all four experiments, humans observe and move more than the rational account prescribes. This gap suggests that observation is driven partly by factors the model omits, such as intrinsic information-seeking or curiosity that is not tied to instrumental payoff \citep{gottlieb2013information, kidd2015psychology}, or a conservative tendency to gather more evidence than a reward-maximizing planner would require. This characterization acts as a next step.

In conclusion, we have argued that previous models of selective social learning fall short because they do not fully exploit Theory of Mind in combination with utility maximization. The Rational Mentalizing model formalizes this idea, thereby gaining the power to learn from sparse, indirect social observation (e.g., from agents with different goals and/or beliefs). Our experiments demonstrate that this model can quantitatively and qualitatively match human behavior in a complex social learning game, taking us one step closer to a more complete theory of human social intelligence.

\section{Methods}

\subsection{Experimental Conditions}

\paragraph{Experiment 1: Single Agent, Single Goal}

Experiment 1 uses a two-agent setting: a red agent (the participant) and a blue expert NPC with full knowledge of amulet locations, who takes the optimal path to its goal. The map contains a single treasure chest. The red agent can observe the blue agent's movements to infer which blue wizard holds the blue amulet.

\paragraph{Experiment 2: Single Agent, Multiple Goals}
Experiment 2 retains the two-agent setting from Experiment 1 but introduces three treasure chests instead of one, adding goal uncertainty. The blue agent's movements now become informative about both amulet locations and the blue agent's intended goal.

\paragraph{Experiment 3: Two Agents, Multiple Goals}
Experiment 3 extends the two-agent setup to a three-agent setting, adding a green agent. The red agent must now decide not only whether to observe but \textit{which} NPC to observe at each timestep---only the selected agent moves during an observation step, while the unselected agent remains still. Both the blue and green agents are experts with full knowledge of their respective amulet locations.

\paragraph{Experiment 4: Expert and Novice Agents, Multiple Goals}
Experiment 4 retains the three-agent, three-chest structure from Experiment 3 but varies the NPCs' experience levels. Each NPC is designated as either an \textbf{expert} or a \textbf{novice}, and the red agent is informed of these designations before each trial. An expert agent has full knowledge of the amulet locations and takes the optimal path to its goal. A novice agent, like the red agent, does not know which wizard holds the amulet. When a novice's goal is blocked by a blue barrier, it approaches the nearest blue wizard, repeating this process until it obtains the blue amulet, and only then proceeds optimally to its goal. Additionally, unlike in previous experiments, notifications about NPC--wizard interactions no longer reveal whether the wizard gave an amulet, further reducing the information available from passive observation. Together, these manipulations allow us to test whether participants calibrate their social learning to the perceived informativeness of each agent.

\subsection{Participants}
All experiments were conducted on Prolific. We recruited 321 US participants across four separate studies: Experiment 1 ($N = 57$; mean age $= 44.7$, $SD = 12.2$; 23 female, 34 male), Experiment 2 ($N = 104$; mean age $= 41.2$, $SD = 11.9$; 41 female, 61 male, 2 non-binary), Experiment 3 ($N = 80$; mean age $= 42.6$, $SD = 13.3$; 37 female, 43 male), and Experiment 4 ($N = 80$; mean age $= 42.4$, $SD = 12.0$; 49 female, 31 male). Participants were paid \$10 per hour, with performance bonuses increasing compensation to \$12 per hour. Participants gave informed consent, and the Harvard University Committee on the Use of Human Subjects approved the experiment.

We applied a participant-level quality filter based on the point system described in the Procedure. Each participant's total score was computed as the sum of unspent points across all non-tutorial levels. Participants whose total score fell below $Q_1 - 1.5 \times \text{IQR}$ were excluded as outliers, where $Q_1$ is the first quartile (computed by linear interpolation) and IQR is the interquartile range of the score distribution within each experiment. This excluded 4 participants from Experiment 1, 10 from Experiment 2, 2 from Experiment 3, and 7 from Experiment 4, yielding final samples of 53, 94, 78, and 73 participants, respectively. Split-half reliability of the observation data was assessed using 1000 random participant splits with Spearman-Brown correction; for the two experiments with multiple observed agents (Experiments 3 and 4), observation counts were stacked across observed agents. This yielded high reliability across all experiments: Experiment 1, $r_{SB} = 0.92$; Experiment 2, $r_{SB} = 0.83$; Experiment 3, $r_{SB} = 0.85$; Experiment 4, $r_{SB} = 0.86$.

\subsection{Procedure}
To mimic the cost structure of real-world social and non-social learning, each action carries a point cost: movement costs 3 points per step, interacting with a wizard costs 5 points, and observing another agent costs 1 point per step. Agents navigate the grid to collect lettered treasure chests, but colored barriers block access to certain chests. To pass a barrier, an agent must first obtain a matching-colored amulet from the correct wizard. Among all wizards of a given color, only one holds the amulet while the rest are decoys. Each trial begins with a fixed point budget; participants keep any unspent points, and their cumulative remaining points across all trials determined a performance bonus (up to \$2 additional compensation). This incentive structure encourages participants to minimize total cost---balancing the cheap but potentially wasteful strategy of prolonged observation against the expensive but direct strategy of independent exploration.

Each participant completed 10 trials presented in one of several pre-randomized orderings, to offset ordering effects. These orderings were also designed so that no two trials shared a similar structure (e.g., none of the maps across the 10 trials were identical). The initial point budget for each level was set by computing the optimal-path cost for the main agent, adding a 20-point buffer, and rounding up to the nearest 5 points.

\subsection{Model Implementation}
The Rational Mentalizing model is implemented in Julia, using the \texttt{PDDL.jl} and \texttt{SymbolicPlanners.jl} libraries for planning and the \texttt{InversePlanning.jl} library with \texttt{Gen.jl} particle filters for Bayesian inference.

\paragraph{Belief Representation}
The observer maintains a belief distribution over which wizard holds the amulet. For each blue wizard, we create a hypothesis state in which that wizard holds the blue amulet. The prior is uniform over all blue wizards. As the observer watches another agent interact with wizards, hypotheses inconsistent with the observed outcomes are pruned, and the belief distribution is updated accordingly.

\paragraph{Inverse Planning}
We pre-compute the posterior distributions over goals and belief states using the Sequential Inverse Plan Search (SIPS) algorithm \citep{zhi2020online} implemented via \texttt{Gen.jl} particle filters. For each map and scenario, we store two probability tables: $P(g \mid a_{1:t}^o)$, the posterior over the observed agent's goal at each timestep, and $P(b \mid a_{1:t}^o, g)$, the posterior over belief states conditioned on each goal hypothesis. A state hypothesis is considered converged when its posterior probability exceeds $0.95$, and goal or state hypotheses with posterior probability below $0.1$ are pruned from consideration. Rather than drawing a fixed number of samples, the filter places one particle on each goal-belief hypothesis, stratified over the candidate goals and belief states, so the discrete hypothesis space is enumerated exhaustively rather than approximated by sampling. All model parameters were set \textit{a priori}, held identical across all four experiments, and never fit to human behavior: the planning temperature was $\beta = 0.5$; the model's per-step action costs matched those faced by participants (movement $3$, wizard interaction $5$, observation $c_{\text{obs}} = 1$ points); and the convergence and pruning thresholds were $0.95$ and $0.1$, respectively.

\paragraph{Self-Exploration Cost}
The cost of non-social learning, $C_{\text{self}}(b_t)$, is estimated by a greedy planner that visits the candidate blue wizards in order of distance from the observer's current position, checking each in turn before continuing to the goal. Writing $w_{(1)}, \dots, w_{(K)}$ for the $K$ wizards consistent with belief $b$ in nearest-first order,
\begin{align*}
    C_{\text{self}}(b) = \sum_{k=1}^{K} \big[\, d(x_{k-1}, w_{(k)}) + c_{\text{int}} \,\big] + d(x_K, x_g),
\end{align*}
where $x_0$ is the observer's current cell, $x_k$ its cell after interacting with $w_{(k)}$, $x_g$ its goal cell, $d(\cdot, \cdot)$ the A$^*$ path cost under a Manhattan heuristic (\texttt{GoalManhattan}), and $c_{\text{int}}$ the interaction cost. The same function, evaluated on the candidate set that remains after observation, supplies the residual planning cost in Eq.~(1): $C_{\text{plan}}^{i} = C_{\text{self}}(b_{T^i})$. Plans are cached by a state-signature key to avoid recomputation.

\paragraph{Observation Utility}
At each timestep, the observer estimates the expected cost of observing each available agent $j$. For each particle $i$ (a goal-belief hypothesis $(g_i, b_i)$) with non-negligible posterior probability, the model looks up the convergence time $T^i$, the number of future observation steps until the belief posterior concentrates on a single wizard, together with the candidate set that remains at convergence. The expected observation cost for agent $j$ is the posterior-weighted average of these per-hypothesis costs,
\begin{align}
    Q_{\text{obs}}^j(t) = \sum_i p_t^{i} \, \big[ c_{\text{obs}} \, T^{i} + C_{\text{plan}}^{i} \big],
\end{align}
identical to Eq.~(1); here each particle $i$ carries weight $p_t^{i} \propto P(g_i \mid a_{1:t}) \, P(b_i \mid a_{1:t}, g_i)$, the joint goal-belief posterior renormalized over the retained hypotheses; $T^{i}$ is the number of remaining observation steps until convergence; and the residual cost $C_{\text{plan}}^{i} = C_{\text{self}}(b_{T^i})$ is the self-exploration cost above, evaluated on the candidates that remain at convergence. Writing $j^* = \arg\min_j Q_{\text{obs}}^j(t)$ for the cheapest agent to watch, the observer then
\begin{align*}
    a_t = \begin{cases} \text{observe } j^*, & \text{if } Q_{\text{obs}}^{j^*}(t) < C_{\text{self}}(b_t),\\[2pt] \text{act}, & \text{otherwise.} \end{cases}
\end{align*}
In the multi-agent setting (Experiments 3--4), the expected costs for all observable agents are computed in parallel.

\subsection{Alternative Model Implementations}

\paragraph{Rational Observer (No Mentalizing)}
This model weighs the cost of self-exploration against the cost of observing for a fixed duration, but does not reason about the other agent's goals or beliefs. It computes two quantities: (1) $C_{\text{self}}(b_t)$, the cost of the observer's greedy exploration over all blue wizards (as above), and (2) $Q_{\text{obs}} = c_{\text{obs}} \cdot T + C_{\text{self}}(b_T)$, where $T$ is the number of steps until the other agent's first wizard interaction and $C_{\text{self}}(b_T)$ is the observer's exploration cost once it gains full information after observing. If $Q_{\text{obs}} < C_{\text{self}}(b_t)$, the agent observes for $T$ steps; otherwise, it skips observation entirely. The decision is binary and made once at the start of each trial.

\paragraph{Mentalizing Observer (No Utility Estimation)}
This model uses Theory of Mind to infer the other agent's goal and belief posteriors but does not evaluate the utility of either social or non-social learning. At each timestep $t$, the observer checks whether, under any plausible goal-state hypothesis, the observed agent's future trajectory could reduce uncertainty about the latent state---specifically, whether observing further would help identify which wizard holds the amulet. The observer compares the current belief state distribution to the predicted distributions under each hypothesis using a distance metric. Formally, the agent continues observing as long as there exists a future horizon over which the state posterior is expected to diverge from the current posterior:
\begin{align}
    d\big(P(b_t^m),\; P(b_t^m \mid a_{t:t+T}^o, g, i)\big) > \epsilon
\end{align}
for at least one $(g, i)$ pair with sufficient probability, where $d(\cdot, \cdot)$ is the Euclidean distance between belief distributions and $\epsilon$ is a threshold. The agent stops observing when the observed agent's trajectory is no longer expected to be informative under any plausible hypothesis---for example, when the observer infers that the agent is pursuing a different goal and its future path will not reveal the amulet location. When two agents are available (Experiments 3--4), the Mentalizing Observer alternates between them, observing each in turn until one yields the needed information or neither is expected to provide further useful observations.

\paragraph{Naive Observer}
The Naive Observer observes the available expert agents until each first interacts with a wizard of the relevant color, at which point it assumes that wizard holds the amulet. This model involves no belief updating, no inference about mental states, and no cost-benefit reasoning. For each expert, we compute its optimal plan and identify the timestep at which it first interacts with a blue wizard, and the Naive Observer observes that agent for exactly that many steps. With a single expert (Experiments 1--2) it observes that agent alone; with two experts (Experiment 3) it observes both, alternating between them until each reaches its first wizard interaction. In Experiment 4, where an expert and a novice NPC are present, the Naive Observer observes only the expert, ignoring the novice entirely.

\bibliographystyle{plainnat}
\bibliography{pnas-sample}

\newpage
\appendix
\renewcommand{\thesection}{Appendix \Alph{section}}
\renewcommand{\thesubsection}{\Alph{section}.\arabic{subsection}}
\renewcommand{\thefigure}{\Alph{section}\arabic{figure}}
\renewcommand{\thetable}{\Alph{section}\arabic{table}}
\renewcommand{\theequation}{\Alph{section}\arabic{equation}}

\section{Supplementary Information}
\setcounter{figure}{0}
\setcounter{table}{0}
\setcounter{equation}{0}
This supplement provides per-experiment breakdowns of the analyses pooled in the main text, the model's reproduction of total and component costs, and the model parameters. All methods are described in the main text.

\subsection{Task workflow}
\begin{figure}[htbp]
\centering
\includegraphics[width=\textwidth]{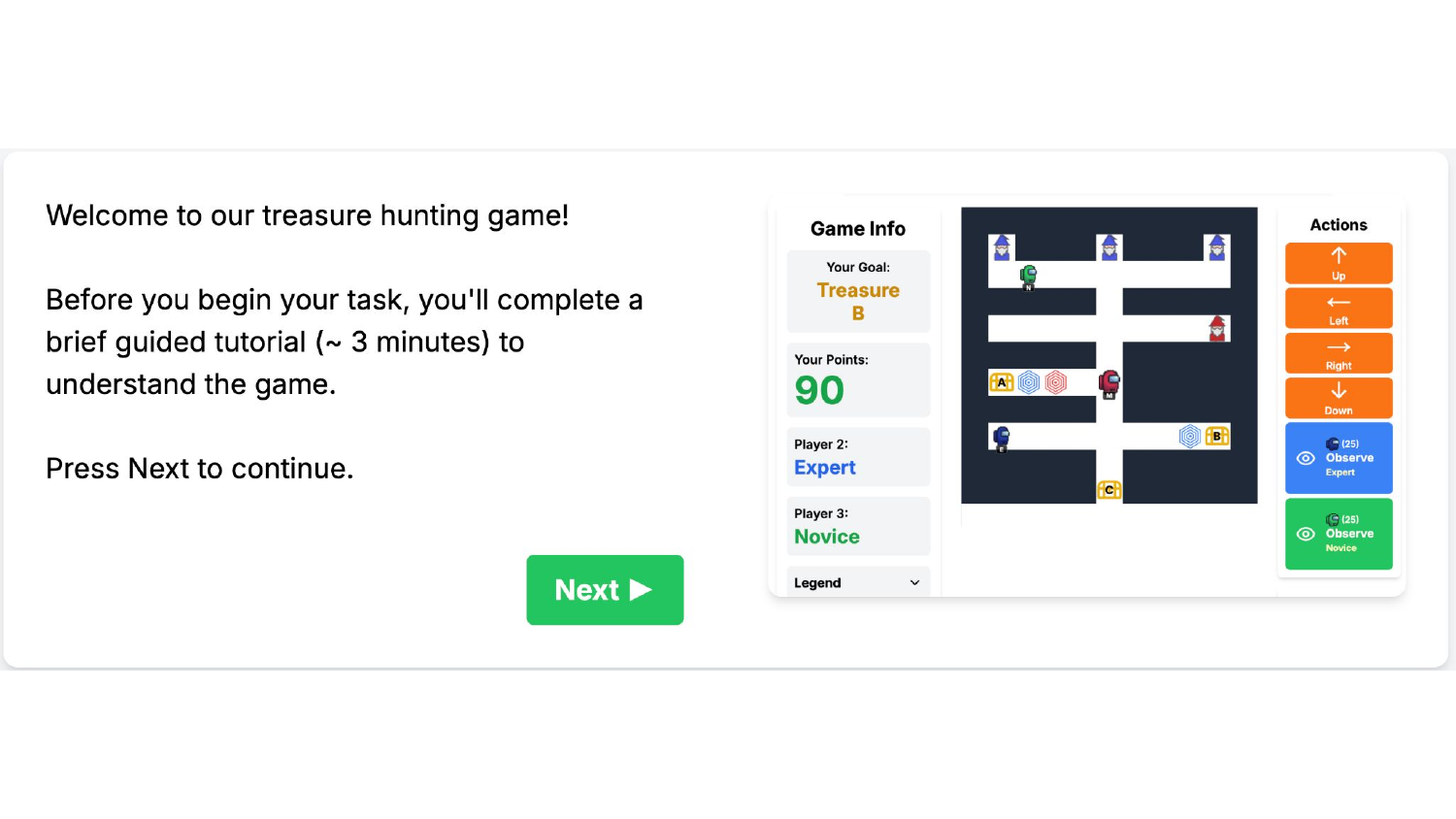}\\[8pt]
\includegraphics[width=\textwidth]{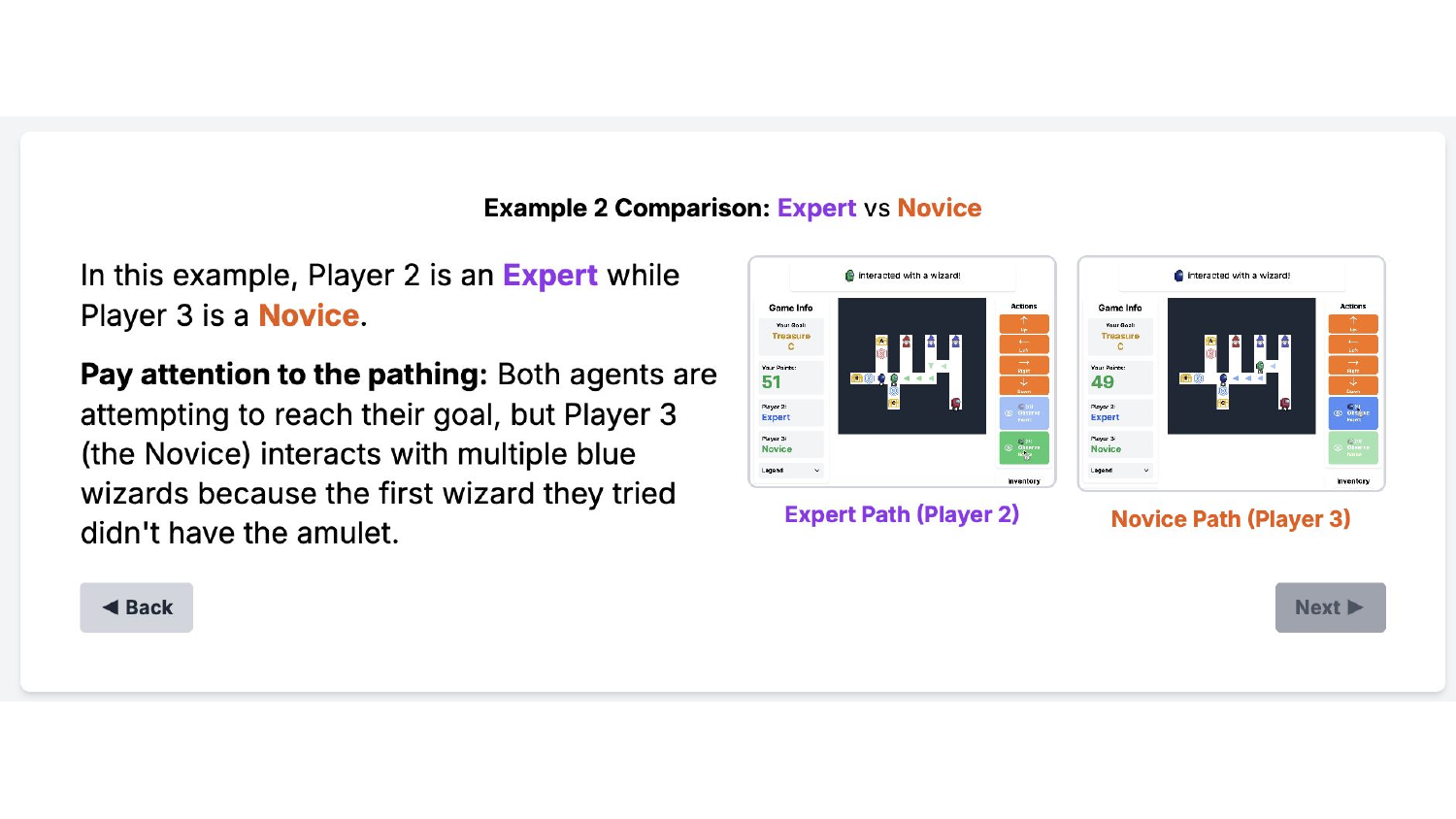}
\caption{\textbf{Task instructions.} \emph{Top:} the welcome screen and game interface---the participant's goal and remaining points, the grid map with the main (red) agent, expert/novice NPCs, wizards, colored barriers, and lettered treasure chests, and the action panel. \emph{Bottom:} a tutorial screen contrasting an expert's and a novice's path to the same goal (the novice checks multiple wizards to find the amulet).}
\label{fig:si_task_instructions}
\end{figure}

\begin{figure}[htbp]
\centering
\includegraphics[width=\textwidth]{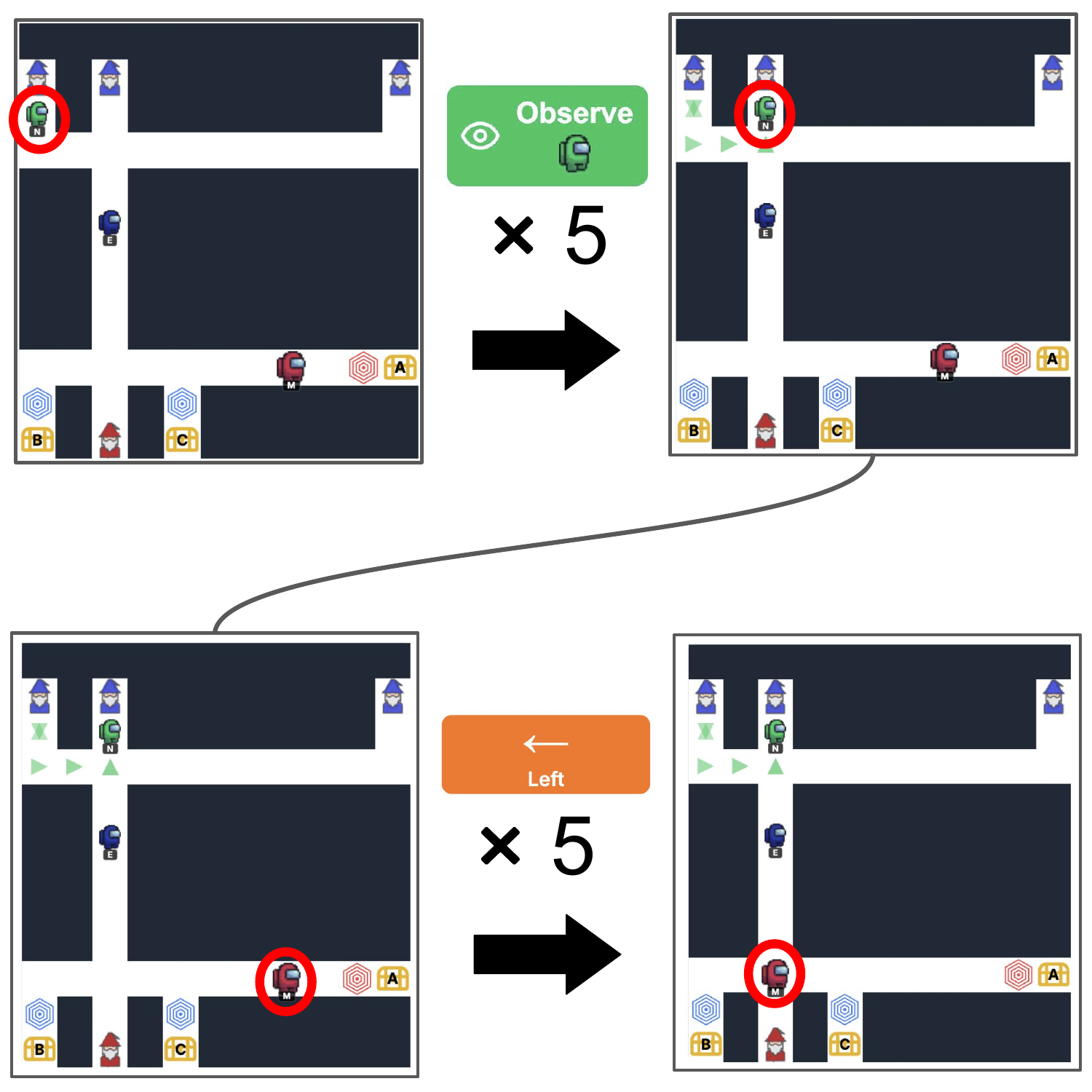}
\caption{\textbf{Example actions.} \emph{Top (observing):} choosing to observe an agent advances only that agent while the participant stays in place; here five observation steps (1 point each) lead the observed agent to interact with a wizard. \emph{Bottom (moving):} choosing a movement action moves the participant's own (red) agent; here five \emph{Left} steps (3 points each).}
\label{fig:si_task_examples}
\end{figure}

Before the main task, participants completed a short guided tutorial. Figure~\ref{fig:si_task_instructions} shows representative instruction screens and the game interface, and Figure~\ref{fig:si_task_examples} illustrates the two action types---observing another agent and moving---each as a before/after state transition.

\begin{figure}[htbp]
\centering
\includegraphics[width=\textwidth]{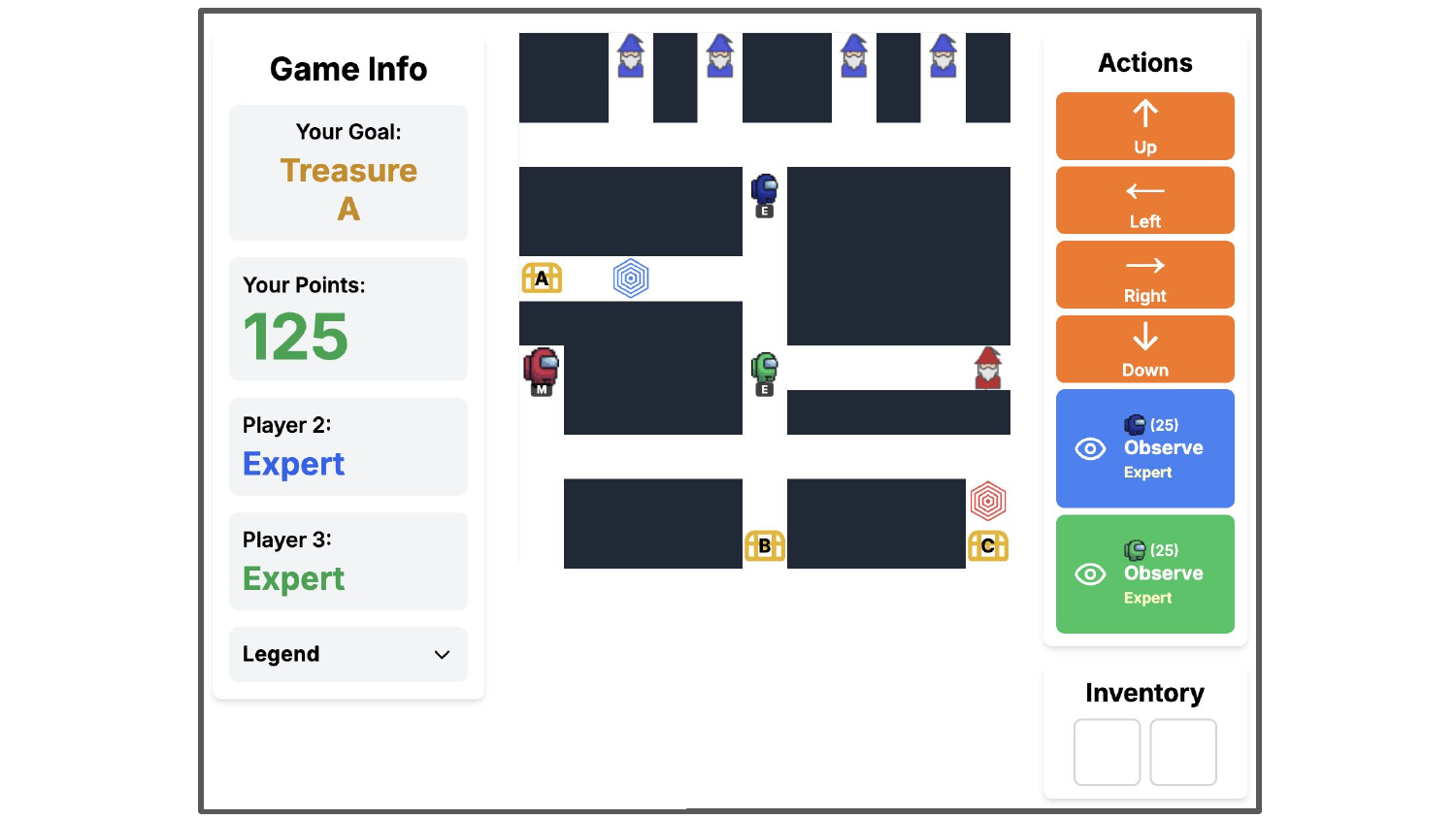}
\caption{\textbf{Game interface.} A single trial as the participant sees it: the \emph{Game Info} panel (current goal, remaining points, and each NPC's expertise), the grid map (red M, the participant; expert/novice NPCs; color-coded wizards; colored barriers; lettered treasure chests), and the \emph{Actions} panel (four movement actions and an \emph{Observe} button for each NPC, with per-step point costs).}
\label{fig:si_task_setup}
\end{figure}

\subsection{Stimuli}
\begin{figure}[htbp]
\centering
\includegraphics[width=\textwidth,height=0.85\textheight,keepaspectratio]{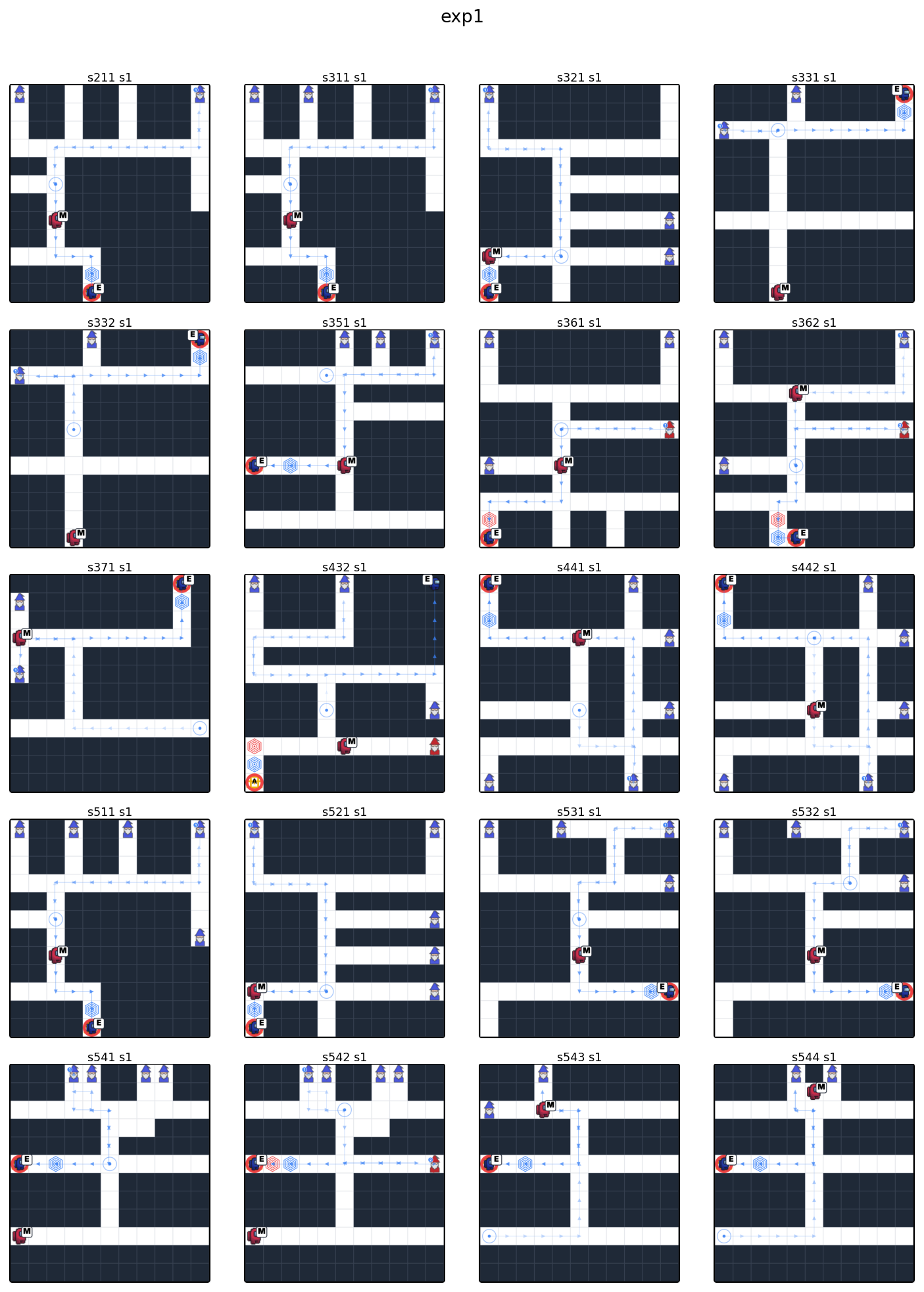}
\caption{\textbf{Experiment 1 stimuli} (single expert, single goal). Every trial map with the expert's reference path. Red M, participant; blue E, expert NPC; arrows, path; hexagons, color-coded amulets; lettered chest, goal. Panels are labelled by map ID.}
\label{fig:si_maps_exp1}
\end{figure}

\begin{figure}[htbp]
\centering
\includegraphics[width=\textwidth,height=0.85\textheight,keepaspectratio]{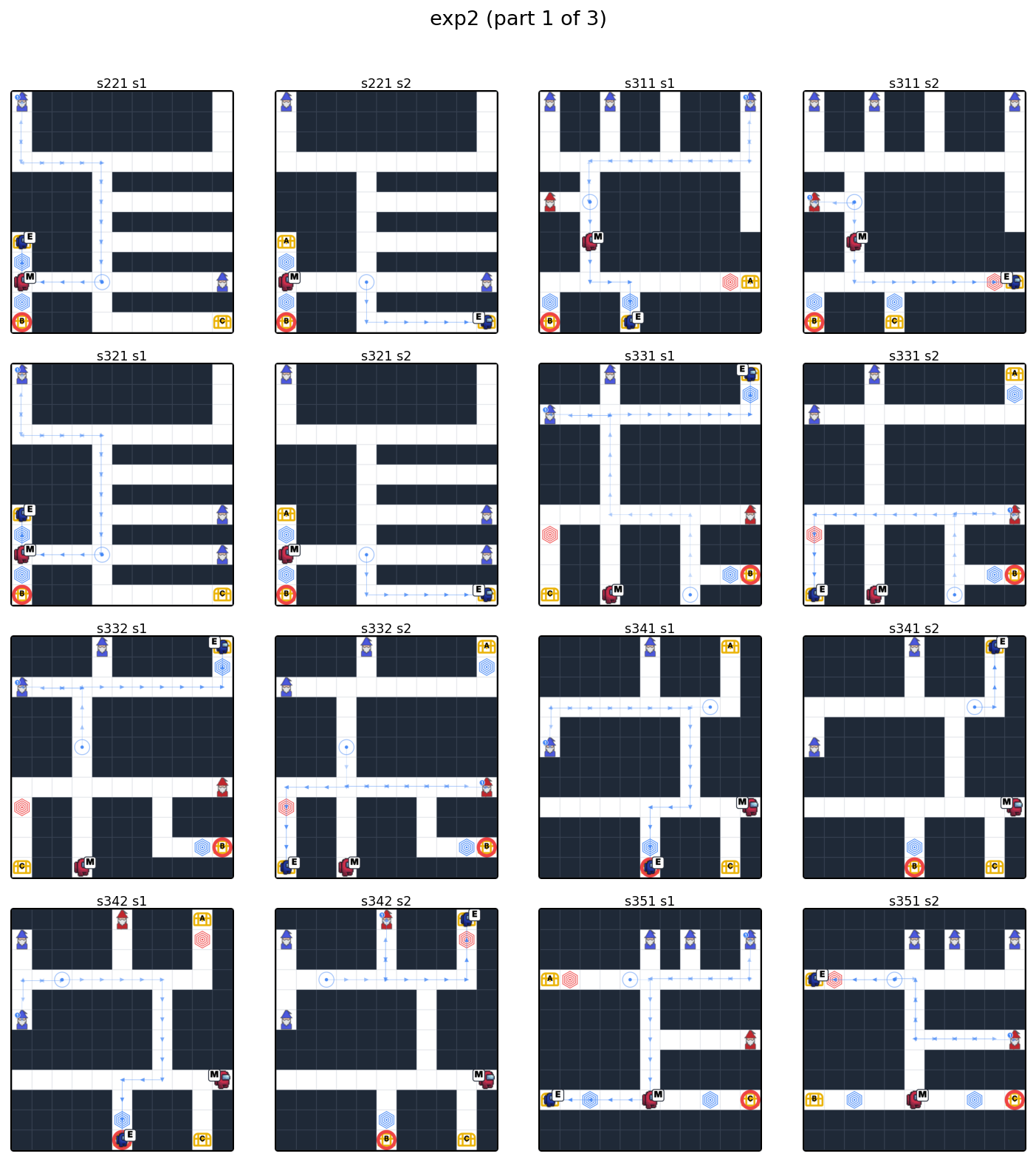}
\caption{\textbf{Experiment 2 stimuli (1 of 3)} (single expert, multiple goals). Every configuration with the expert's reference path; scenario pairs (\texttt{s1}/\texttt{s2}) are adjacent and differ in the expert's goal.}
\label{fig:si_maps_exp2a}
\end{figure}

\begin{figure}[htbp]
\centering
\includegraphics[width=\textwidth,height=0.85\textheight,keepaspectratio]{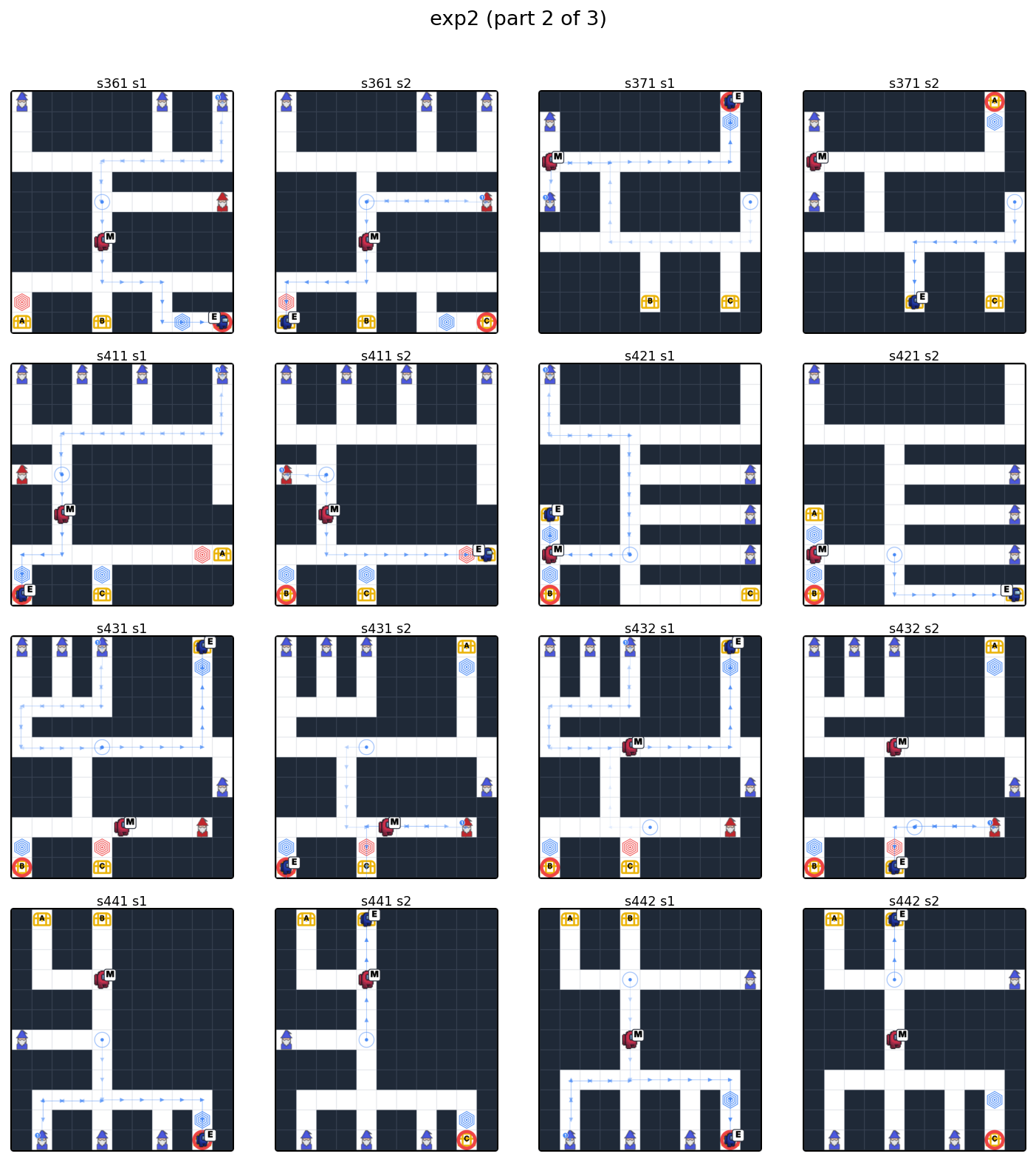}
\caption{\textbf{Experiment 2 stimuli (2 of 3).} Continued from Fig.~\ref{fig:si_maps_exp2a}.}
\label{fig:si_maps_exp2b}
\end{figure}

\begin{figure}[htbp]
\centering
\includegraphics[width=\textwidth,height=0.85\textheight,keepaspectratio]{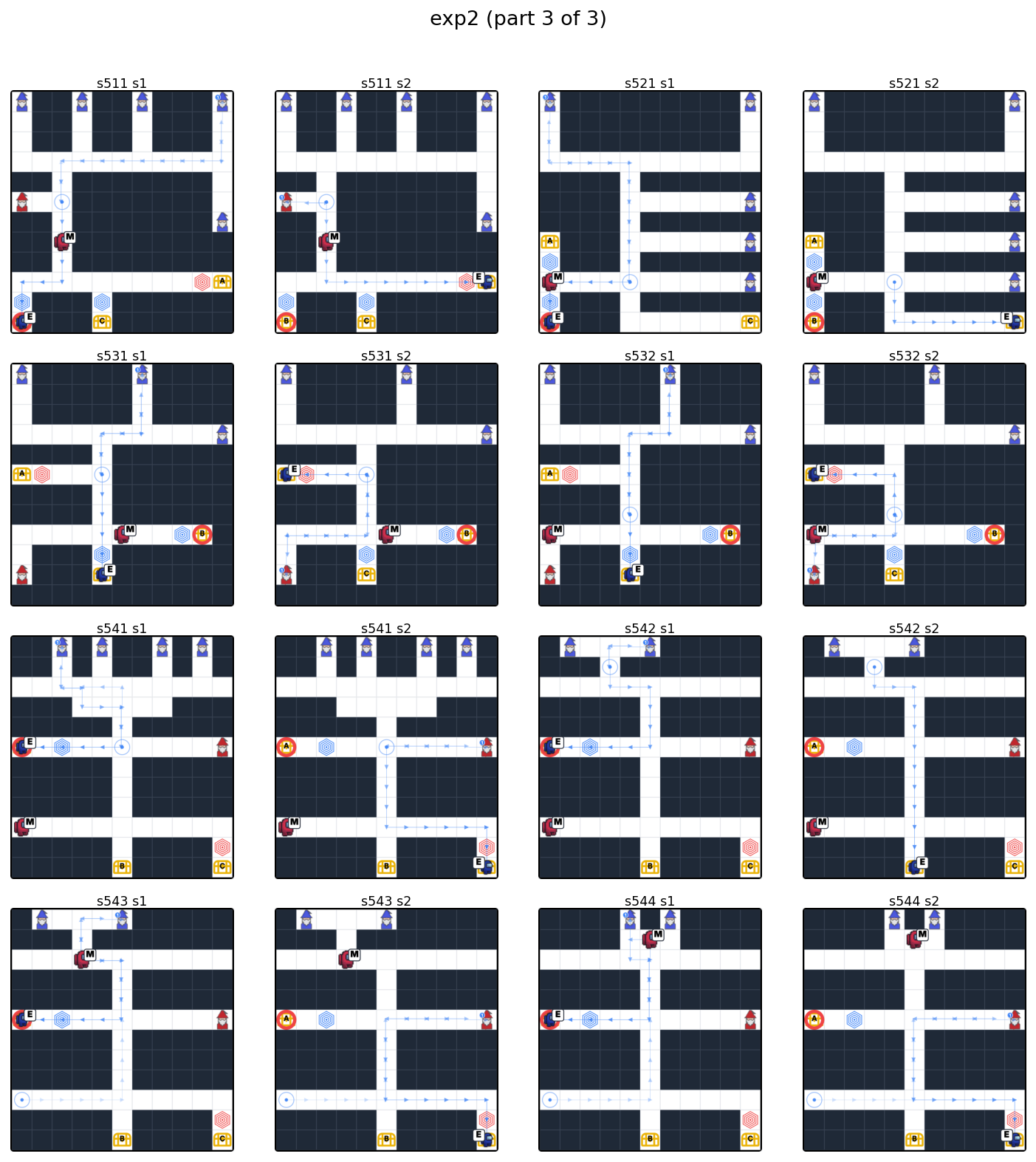}
\caption{\textbf{Experiment 2 stimuli (3 of 3).} Continued from Fig.~\ref{fig:si_maps_exp2a}.}
\label{fig:si_maps_exp2c}
\end{figure}

\begin{figure}[htbp]
\centering
\includegraphics[width=\textwidth,height=0.85\textheight,keepaspectratio]{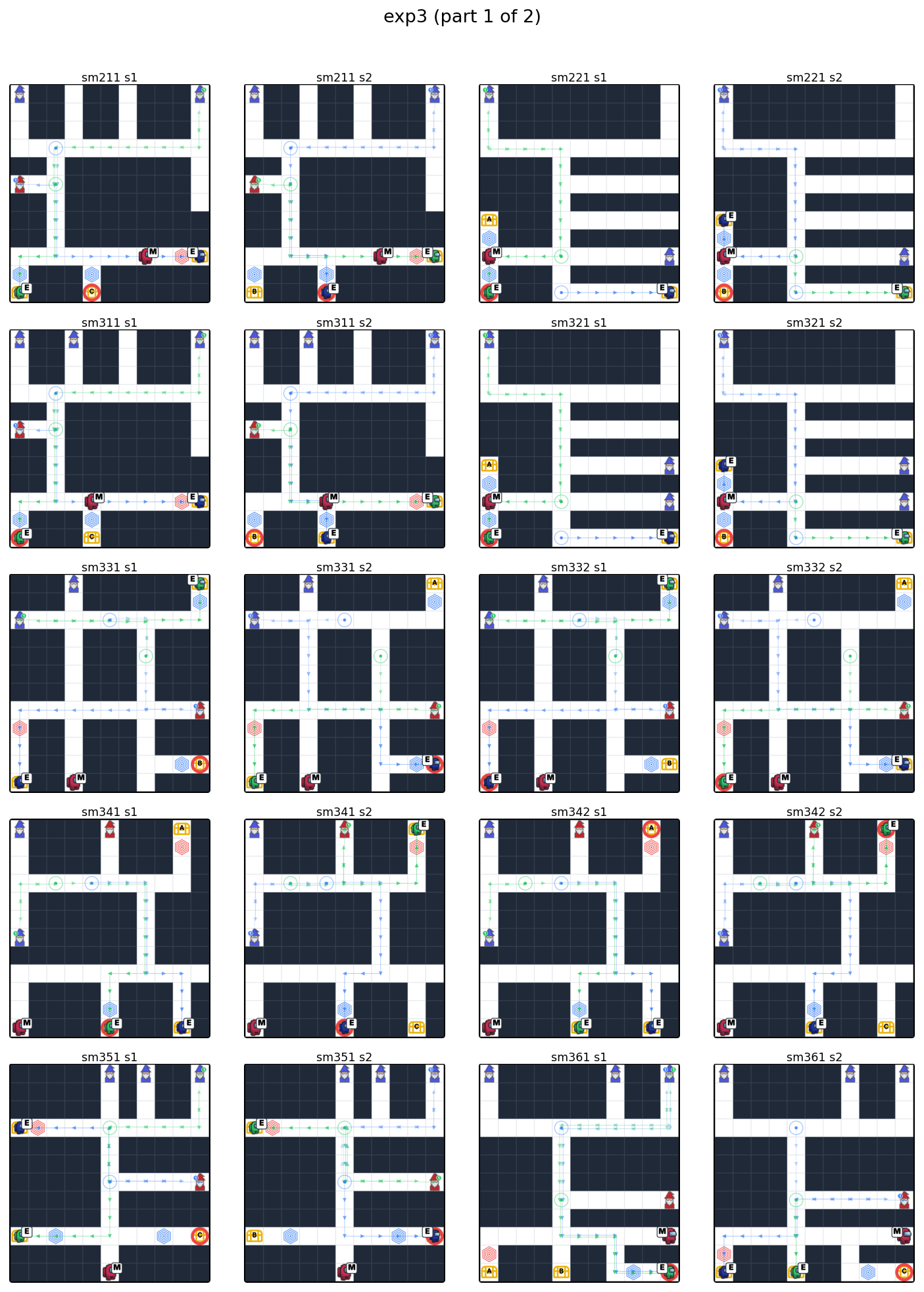}
\caption{\textbf{Experiment 3 stimuli (1 of 2)} (two experts, multiple goals). Every configuration with both NPCs' reference paths (blue and green); scenario pairs (\texttt{s1}/\texttt{s2}) are adjacent.}
\label{fig:si_maps_exp3a}
\end{figure}

\begin{figure}[htbp]
\centering
\includegraphics[width=\textwidth,height=0.85\textheight,keepaspectratio]{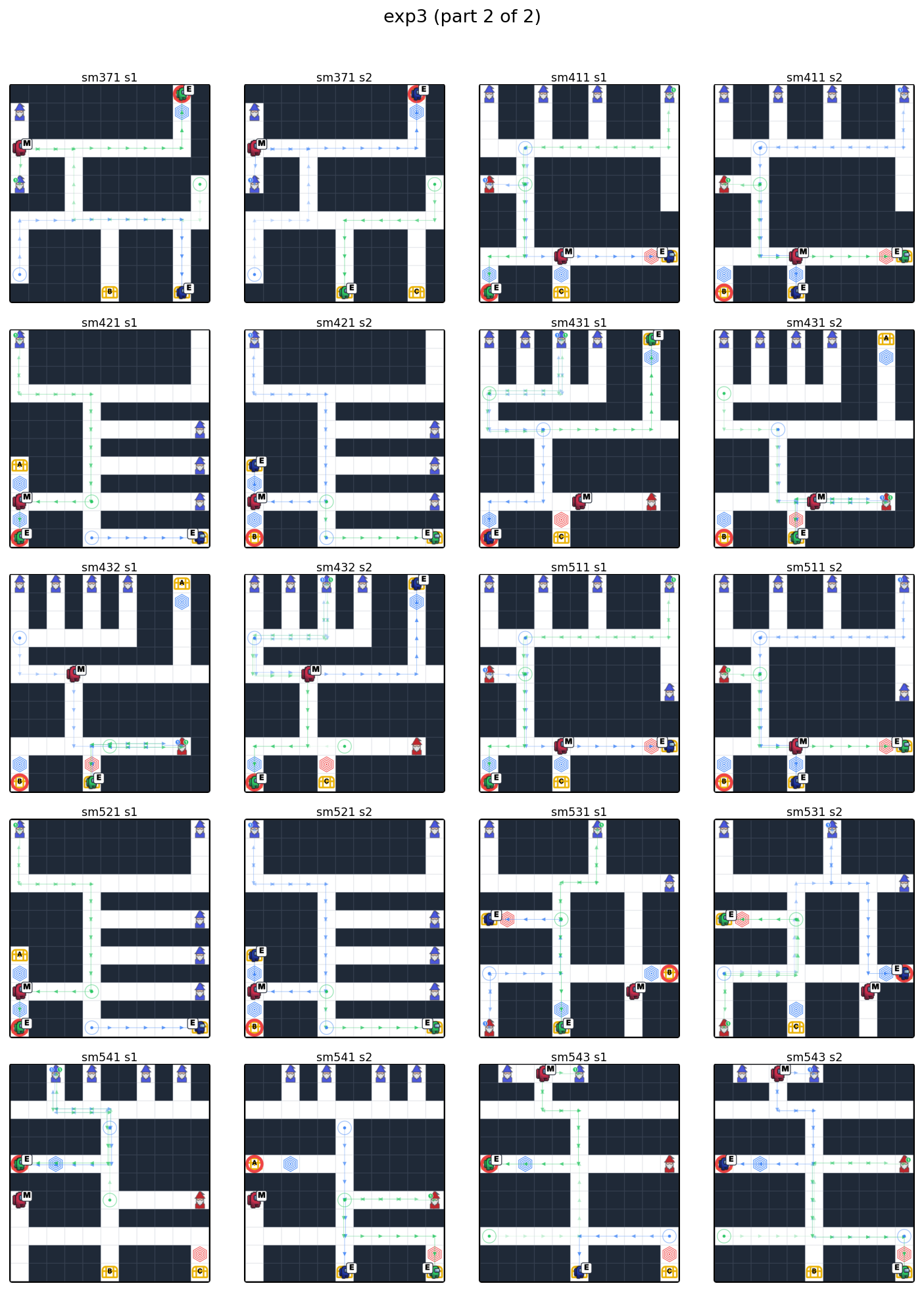}
\caption{\textbf{Experiment 3 stimuli (2 of 2).} Continued from Fig.~\ref{fig:si_maps_exp3a}.}
\label{fig:si_maps_exp3b}
\end{figure}

\begin{figure}[htbp]
\centering
\includegraphics[width=\textwidth,height=0.85\textheight,keepaspectratio]{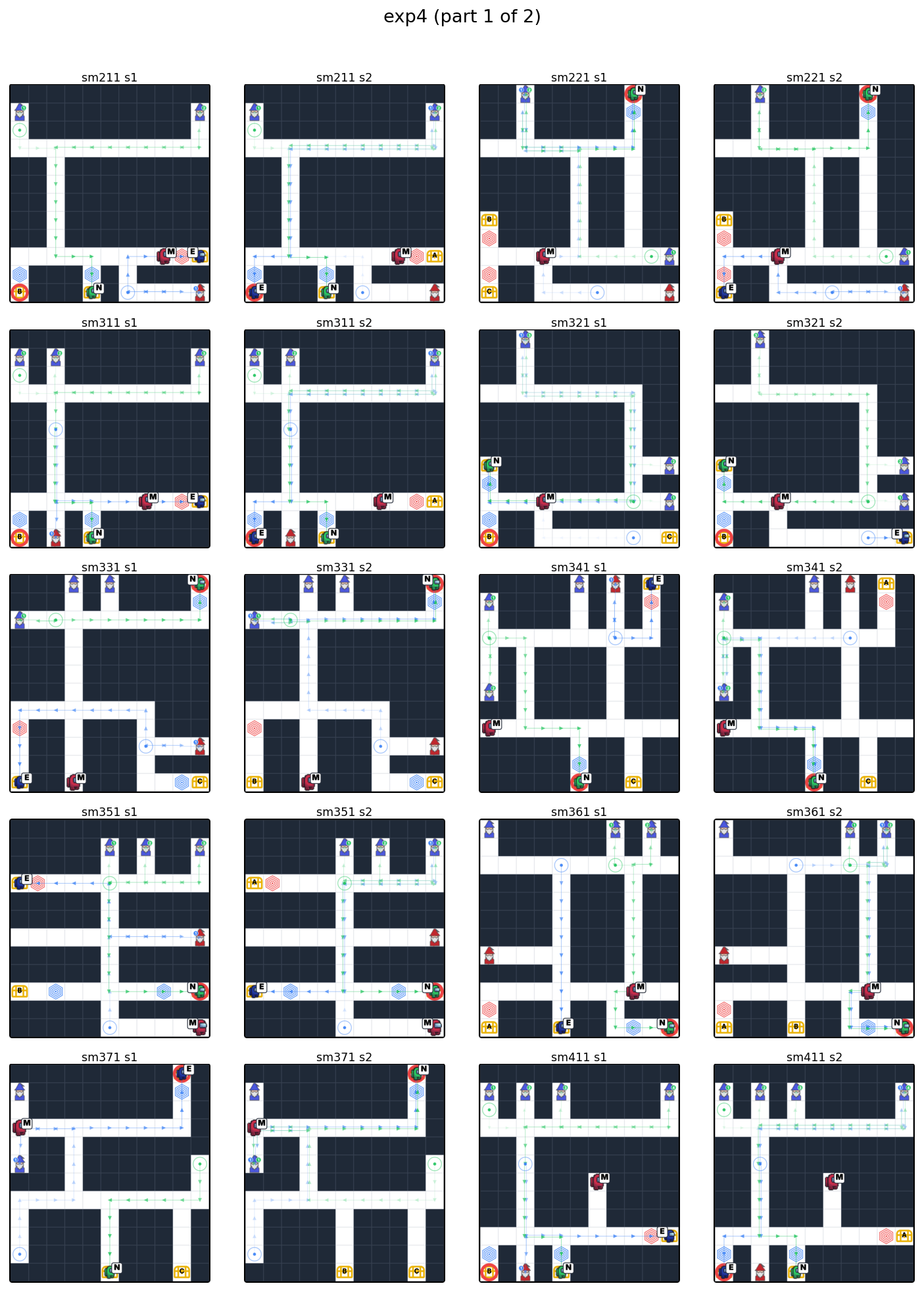}
\caption{\textbf{Experiment 4 stimuli (1 of 2)} (expert and novice, multiple goals). Every configuration with the expert's (blue) and novice's (green) reference paths; the novice's path may check several wizards before finding the amulet. Scenario pairs (\texttt{s1}/\texttt{s2}) are adjacent.}
\label{fig:si_maps_exp4a}
\end{figure}

\begin{figure}[htbp]
\centering
\includegraphics[width=\textwidth,height=0.85\textheight,keepaspectratio]{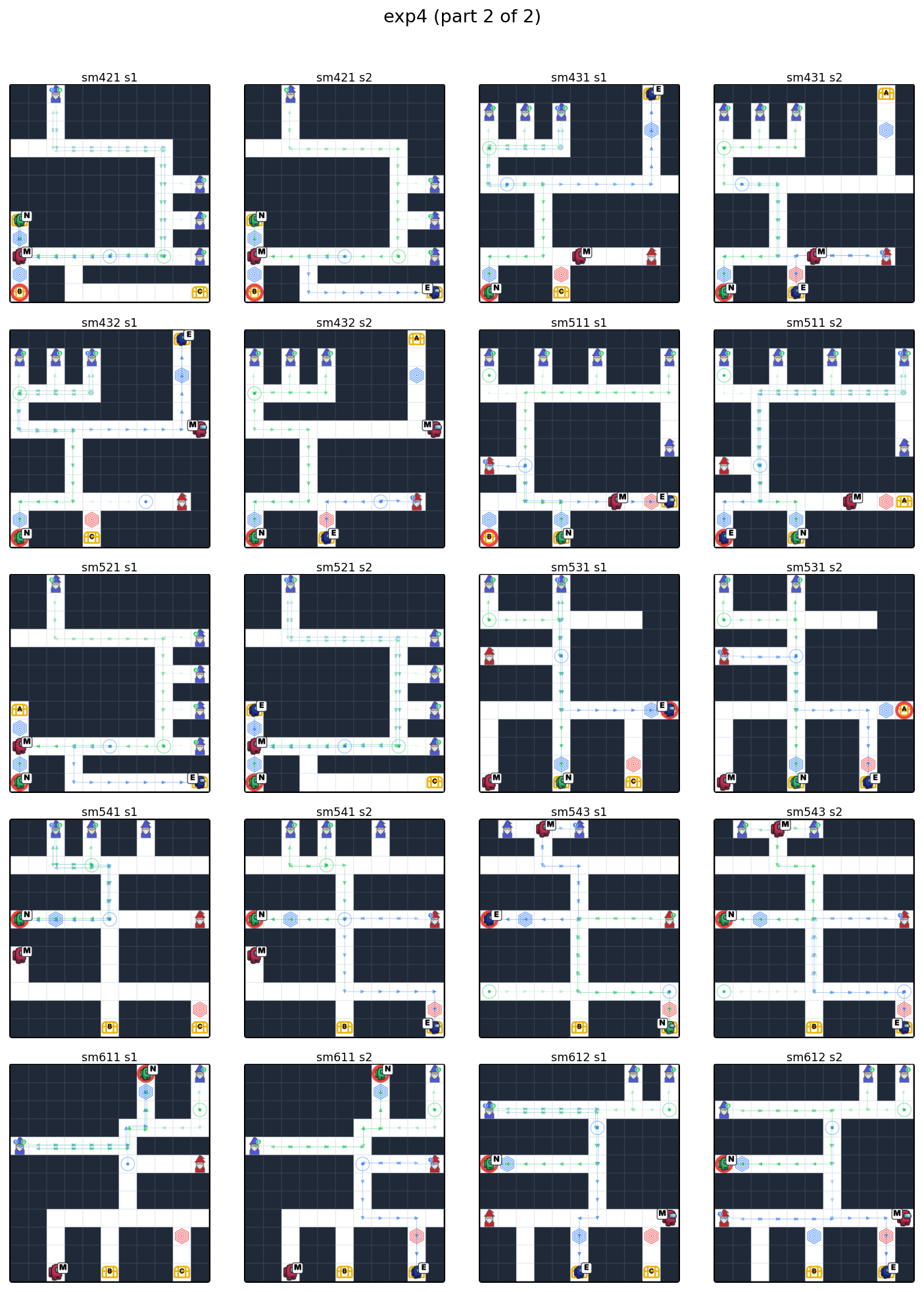}
\caption{\textbf{Experiment 4 stimuli (2 of 2).} Continued from Fig.~\ref{fig:si_maps_exp4a}.}
\label{fig:si_maps_exp4b}
\end{figure}

Each experiment used a pool of grid maps (all $12\times11$ cells); each participant completed 10 trials drawn from the pool. A base map can appear in two \emph{scenarios} that share its layout but differ in the NPC goal assignments (and, in Experiment~4, agent expertise), which changes what each agent's path reveals. Table~\ref{tab:si_stimuli} summarizes the set, and Figures~\ref{fig:si_maps_exp1}--\ref{fig:si_maps_exp4b} show every configuration with its agents' reference paths (scenario pairs are shown adjacently, labelled \texttt{s1}/\texttt{s2}). All maps and reference paths are also available in the repository.

\begin{figure}[htbp]
\centering
\includegraphics[width=\textwidth]{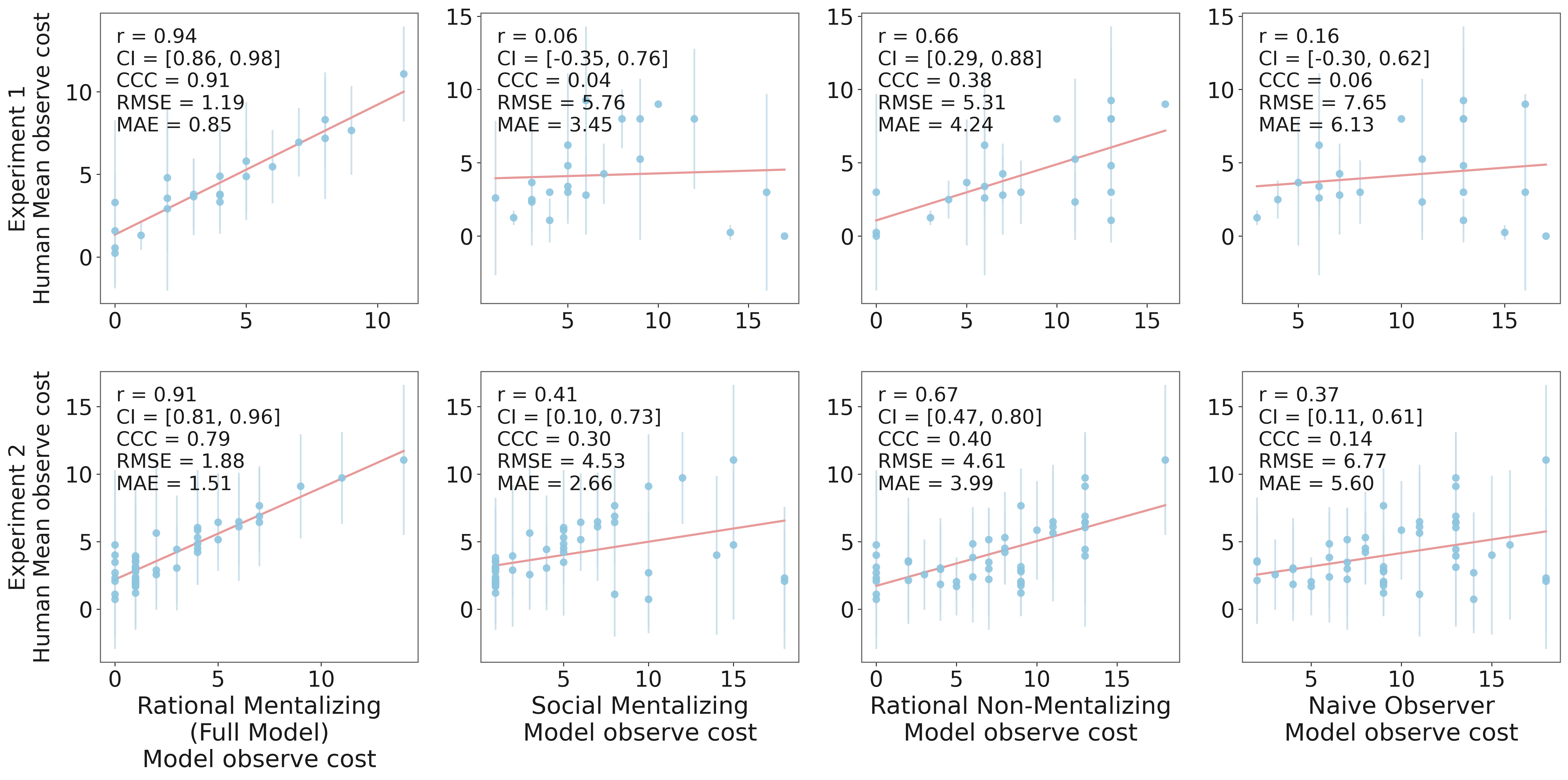}
\caption{\textbf{Observation cost, Experiments 1--2} (single observed agent), human vs.\ model. Rows: Experiment 1; Experiment 2. Columns: the four models, each with Pearson $r$, 95\% bootstrap CI, Lin's CCC, RMSE, and MAE.}
\label{fig:si_cost_obs_12}
\end{figure}

\begin{figure}[htbp]
\centering
\includegraphics[width=\textwidth,height=0.85\textheight,keepaspectratio]{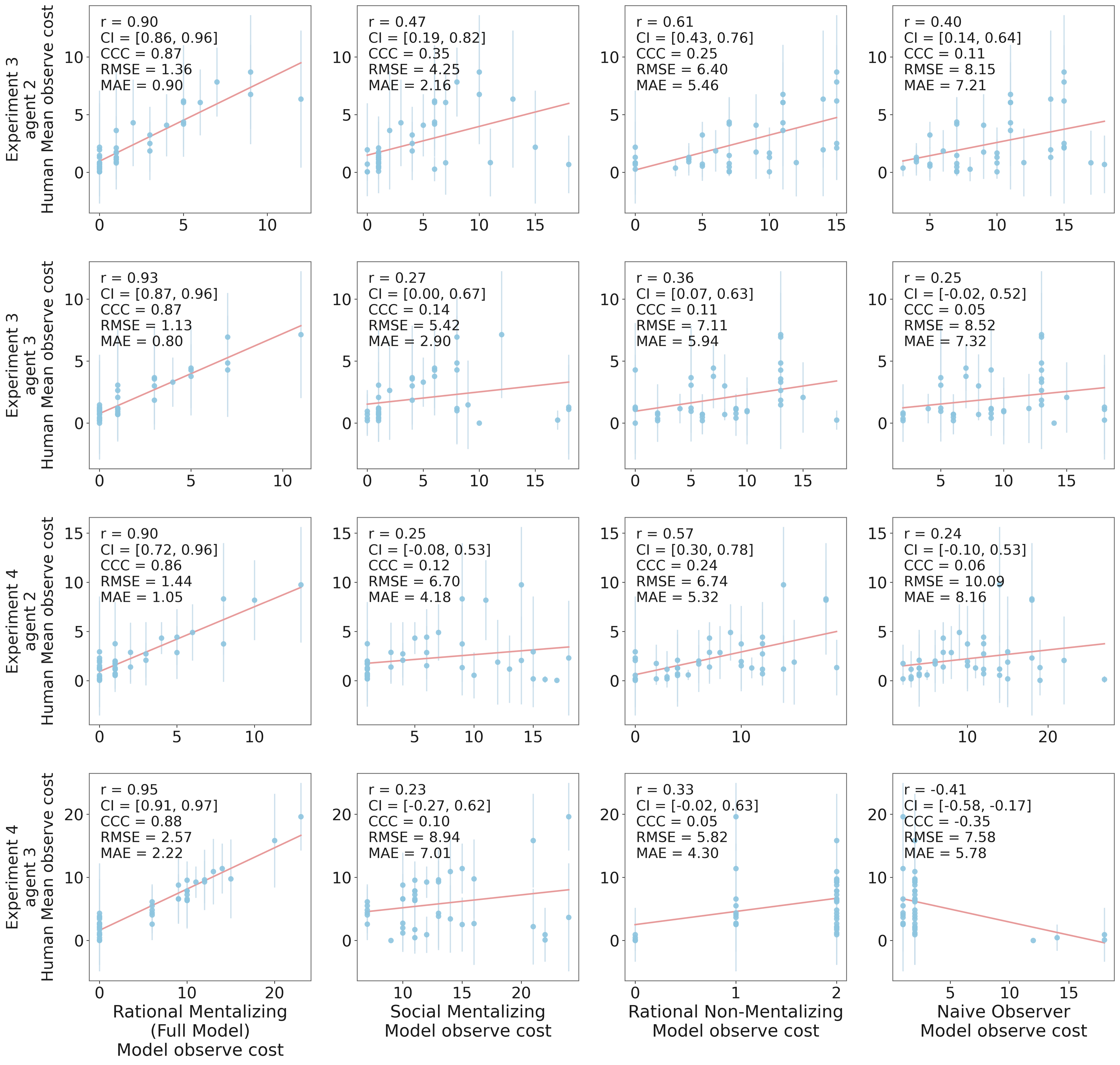}
\caption{\textbf{Observation cost, Experiments 3--4}, split by observed agent. Rows: Experiment 3 (agent 2); Experiment 3 (agent 3); Experiment 4 (agent 2); Experiment 4 (agent 3). Columns and statistics as in Fig.~\ref{fig:si_cost_obs_12}. Unlike total cost (Fig.~\ref{fig:si_cost_total}), observation cost separates the models: only the Rational Mentalizing model matches it.}
\label{fig:si_cost_obs_34}
\end{figure}

\begin{figure}[htbp]
\centering
\includegraphics[width=\textwidth]{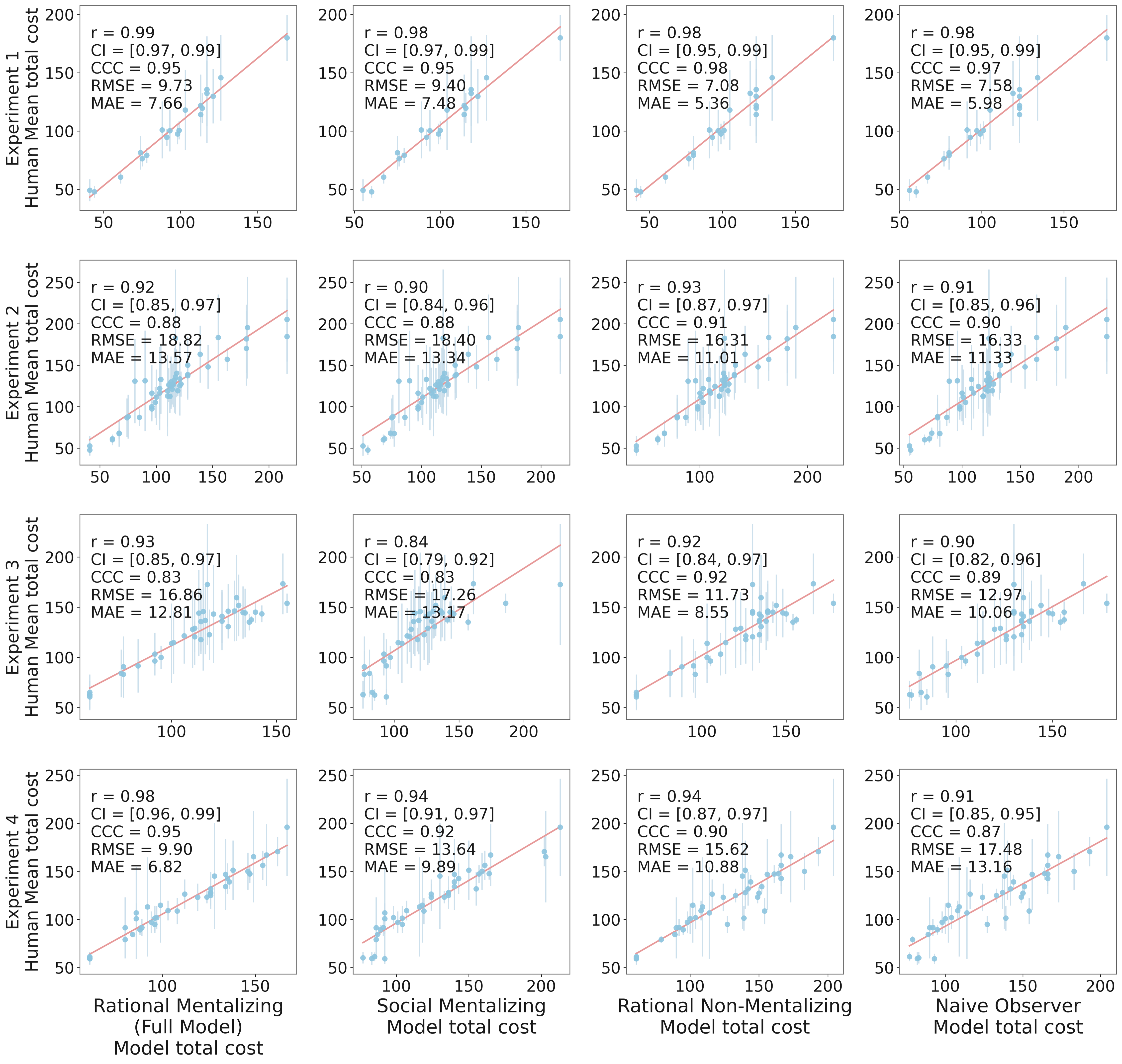}
\caption{\textbf{Total cost}, human vs.\ model, by experiment (rows) and model (columns), with Pearson $r$, Lin's concordance correlation coefficient (CCC), RMSE, and MAE. All four models reproduce total cost well, showing that overall cost is not diagnostic for distinguishing them.}
\label{fig:si_cost_total}
\end{figure}

\begin{figure}[htbp]
\centering
\includegraphics[width=\textwidth]{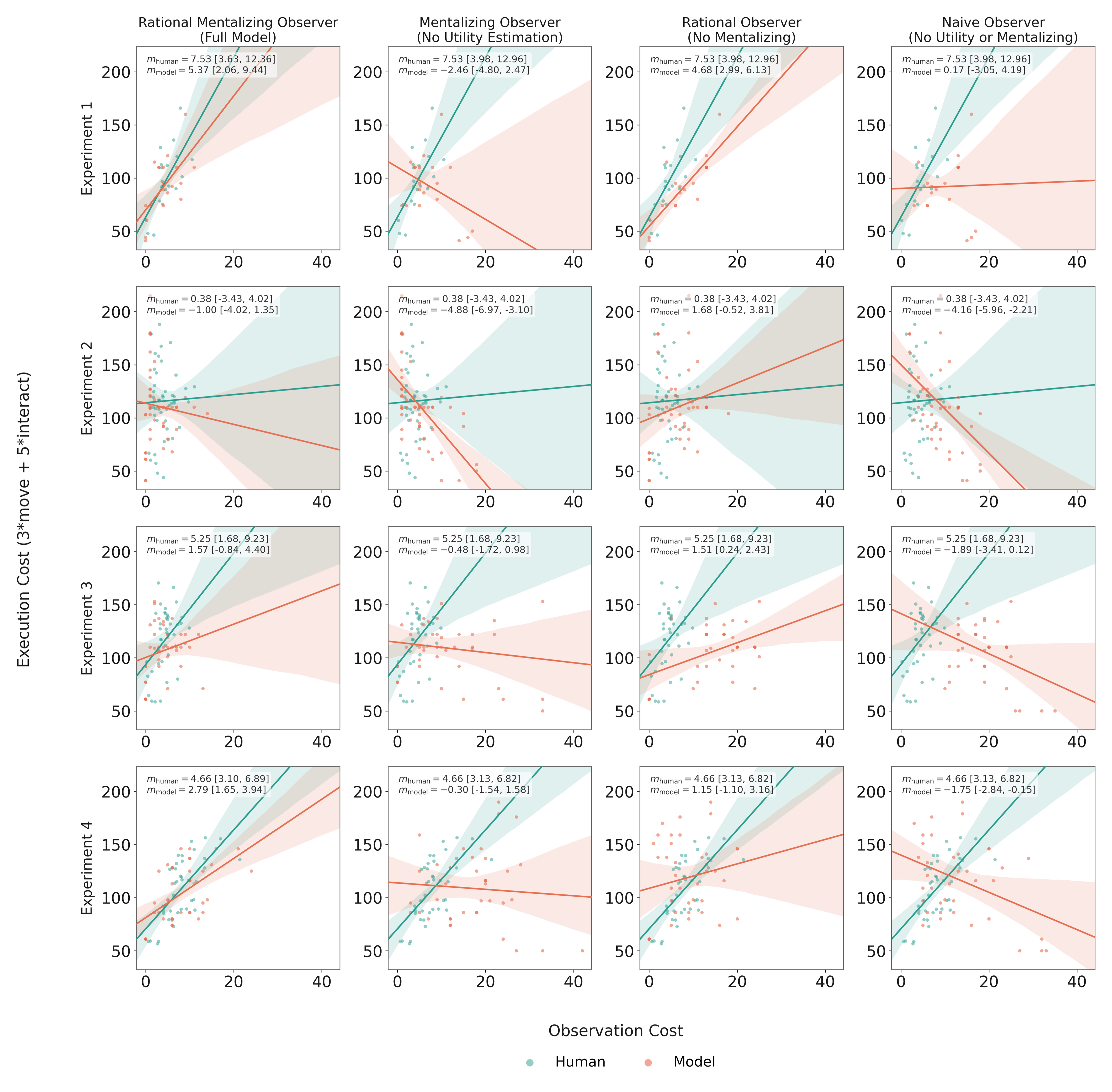}
\caption{\textbf{Observation--execution trade-off by experiment} (rows) and model (columns). Teal: human; coral: model; lines are fitted slopes. The positive human slope, and the full model's match to it, hold within every experiment.}
\label{fig:si_tradeoff_facet}
\end{figure}

\begin{table}[htbp]\centering
\caption{Stimulus inventory. All maps are $12\times11$ grids; each participant completed 10 trials drawn from the experiment's pool. ``Scenarios'' share a base map's layout but differ in NPC goal assignments (and, in Experiment~4, agent expertise). Counts are the configurations entering the analyses (shown in full in Figs.~\ref{fig:si_maps_exp1}--\ref{fig:si_maps_exp4b}).}
\label{tab:si_stimuli}
\begin{tabular}{lccccc}
\toprule
Experiment & Base maps & Scenarios/map & Configs & NPC agents & Goal chests \\
\midrule
1 (single agent, single goal)       & 20 & 1 & 20 & 1 expert            & 1 \\
2 (single agent, multiple goals)    & 24 & 2 & 48 & 1 expert            & 3 \\
3 (two agents, multiple goals)      & 20 & 2 & 40 & 2 experts           & 3 \\
4 (expert + novice, multiple goals) & 20 & 2 & 40 & 1 expert + 1 novice & 3 \\
\midrule
Total                               & 84 & --- & 148 & & \\
\bottomrule
\end{tabular}

\vspace{2.5em}

\caption{Model parameters. All values were set \emph{a priori} and shared across Experiments 1--4; none were fit to human data.}
\label{tab:si_params}
\begin{tabular}{lll}
\toprule
Parameter & Symbol & Value \\
\midrule
Planning (Boltzmann) temperature & $\beta$ & $0.5$ \\
Movement cost & & $3$ points \\
Wizard-interaction cost & $c_{\text{int}}$ & $5$ points \\
Observation cost & $c_{\text{obs}}$ & $1$ point \\
State-convergence threshold & & $0.95$ \\
Hypothesis-pruning threshold & & $0.1$ \\
Belief-divergence threshold (Mentalizing Observer) & $\epsilon$ & $0.1$ \\
Goal--belief hypotheses & & one particle per $(g,b)$ pair (exhaustive) \\
Self-exploration planner & & A$^{*}$ with Manhattan heuristic \\
\bottomrule
\end{tabular}
\end{table}

\FloatBarrier

\subsection{Data and Code Availability}
The behavioral game, the Rational Mentalizing model and the three ablation implementations, the anonymized human data, and all analysis and plotting code are available at \url{https://osf.io/e93kh/}. The model is implemented in Julia using the \texttt{PDDL.jl}, \texttt{SymbolicPlanners.jl}, and \texttt{InversePlanning.jl} libraries with \texttt{Gen.jl} particle filters; analyses and figures are produced with the Python scripts in the same repository.

\end{document}